\documentclass[fleqn,preprint,review,11pt]{elsarticle}
\usepackage{graphicx}
\usepackage{epstopdf}
\usepackage{physics}
\usepackage{amssymb}
\usepackage{amsmath}
\usepackage{amsfonts}
\usepackage[shortlabels]{enumitem}
\usepackage{mathrsfs}
\usepackage{natbib}
\usepackage{xcolor}
\usepackage[draft]{changes}
\usepackage{algorithm}
\usepackage[noend]{algpseudocode}
\usepackage{float}
\usepackage{subcaption}
\usepackage{lineno}

\makeatletter
\let\LN@align\align
\let\LN@endalign\endalign
\renewcommand{\align}{\linenomath\LN@align}
\renewcommand{\endalign}{\LN@endalign\endlinenomath}
\let\LN@gather\gather
\let\LN@endgather\endgather
\renewcommand{\gather}{\linenomath\LN@gather}
\renewcommand{\endgather}{\LN@endgather\endlinenomath}
\makeatother

\newcommand{\noah}[1]{\noindent \color{black} #1\normalcolor}

\journal{Journal of Computational Physics}

\begin{document}

\begin{frontmatter}



\title{A Level-Set Immersed Boundary Method for  Reactive Transport in Complex Topologies with Moving Interfaces}

\author[1]{Mehrdad Yousefzadeh\corref{contrib}}
\author[1]{Yinuo Yao\corref{contrib}}
\author[1]{Ilenia Battiato\corref{corauthor}}
\address[1]{Departmenet of Energy Science and Engineering, Stanford University, Stanford, CA 94305, USA}

\cortext[corauthor]{Corresponding author.}
\cortext[contrib]{Both authors contributed equally to this work.}

\begin{abstract}
A simulation framework based on the level-set and the immersed boundary methods (LS-IBM) has been developed for reactive transport problems in porous media involving a moving solid-fluid interface. The interface movement due to  surface reactions is tracked by the level-set method, while the immersed boundary method captures the momentum and mass transport at the interface. The proposed method is capable of accurately modeling transport  near evolving boundaries in Cartesian grids. The framework formulation guarantees second order accuracy {in space}. Since the interface velocity is only defined at the moving boundary, an interface velocity propagation method is also proposed. The  method can be applied to other moving interface  problems of the ``Stefan'' type. Here, we validate the proposed LS-IBM both for flow and transport close to an immersed object with reactive boundaries as well as for crystal growth. The proposed method provides a powerful tool to model more realistic problems involving moving reactive interfaces in complex domains. 
\end{abstract}


\end{frontmatter}


\section{Introduction}
Dissolution and/or precipitation of  solid minerals due to heterogeneous reactions occurring at solid-fluid interfaces is important in a variety of geological \cite{morse2007calcium, lichtner2018reactive, yousefzadeh2017physics}  and engineered \cite{ryan2013computational, tan2016computational} porous media. The dissolution of  solid rock matrix due to acid injection into  fractures and near wellbores leads to dramatic localized  increases in permeability \cite{pournik2014effect, teklu2017experimental, pournik2019productivity, deng2017alteration}. During CO$_{2}$ sequestration into deep reservoirs, supercritical CO$_{2}$  reacts with water to form carbonic acid and leads to the dissolution of the carbonate host rock \cite{rathnaweera2016experimental, xiao2009effects, xu2003reactive, huerta2016reactive}. Mineral precipitation is a vital process in the immobilization of heavy metals in contaminated soils and water resources \cite{wuana2011heavy, guo2018stability}.  Dendrite growth on the electrode surface of a battery is another example of precipitation in reactive systems \cite{ryan2013computational, tan2016computational}. 

Multiple approaches have been developed to numerically model reactive transport in complex porous structures. These approaches include direct numerical simulations (DNS), pore network models (PNM),  micro-continuum models based on the coupling of Darcy-Brinkman-Stokes equations (DBS) \cite{soulaine2017mineral}, smoothed particle hydrodynamics (SPH) \cite{tartakovsky2007smoothed}, lattice Boltzmann method (LBM) \cite{huber2014new, parmigiani2011pore, prasianakis2017deciphering} and immersed boundary methods (IBM) \cite{yousefzadeh2019high,Yao2022-aq}. The presence of precipitation and/or dissolution at fluid-solid interfaces poses additional challenges, since mass and momentum transfer are coupled through a moving boundary. Such a class of problems involving moving boundaries are generally referred to as ``Stefan problems" and require  an  explicit treatment of the interface separating  solid and fluid. 

Numerical methods for problems involving  interface phenomena are broadly categorized in two classes:  sharp interface and diffusive interface methods. On the one hand, diffuse interface methods (e.g. the phase field method) smooth out any  jump discontinuity across the interface over a range of grid cells (i.e. the support or width of the smoothed delta function). Diffuse interface methods are constructed such that in the limit of an infinitesimal interface width (the so-called `sharp interface limit'), the correct interfacial dynamics is recovered. On the other hand,  sharp interface methods (e.g. VOF, IBM) preserve any sharp discontinuity  in the variables  across the interface. Generally, sharp interface methods are preferable over diffuse interface methods for interface problems: when  mass, heat or flow boundary layers are thin,  diffusive interface methods may encounter some difficulties, since the approximate width of the interface might be comparable to the boundary layer thickness. This would require very fine grids in proximity of the interface and will result in increased computational burden.

A number of methods have been proposed to handle moving boundaries \cite{molins2015reactive}. In the front tracking method (FTM), introduced by Unverdi and Tryggvason \cite{unverdi1992front}, the interface is represented explicitly. Juric and  Tryggvason \cite{juric1996front} used FTM to model the solid boundary evolution in a dendrite solidification problem. Implicit models of the interface including the level set method (LSM) \cite{osher1988fronts, osher2001level}, the volume of fluid (VOF) \cite{hirt1981volume} and the phase field method (PFM)  \cite{xu2011phase, xu2012phase} are  preferable over explicit front tracking methods,  since they are robust in handling  complex topological changes.  
Xu \emph{et al.} \cite{xu2012phase} used the phase field method to model the dynamics of a moving interface due to a surface chemical reaction. In the level set method (LSM), first introduced by Osher and Sethian \cite{osher1988fronts} for tracking the front propagation of a curvature dependent interface, the interface is represented as the zero contour of a higher dimensional function. LSM has been extensively used to model  interfaces on static Cartesian grids and is used in both diffusive and sharp interface approaches, depending on  how  boundary conditions are enforced.  In the context of fluid dynamic modeling, the LSM is primarily used for tracking  fluid-fluid interfaces in multiphase flow problems \cite{osher2004level, fedkiw1999non}. Li \emph{et al.} \cite{li2008level, li2010three} proposed a level set method for precipitation and dissolution problems. Since the  works by Li \emph{et al.}, the LSM has been broadly used for reactive transport problems involving  moving interfaces \cite{molins2017mineralogical, molins2020simulation, chai2019finite}.  
 
 Once the interface is properly captured, the boundary conditions associated with the interaction between the solid and fluid domains need to be imposed. The irregularity of the solid geometry imposes additional difficulties for  Cartesian grids. Various Cartesian grid-based algorithms have been suggested for reactive type boundary conditions, including embedded boundary method (EBM) \cite{trebotich2015adaptive, trebotich2014high} and immersed boundary methods (IBM) \cite{yousefzadeh2019high, luo2016ghost}. Immersed boundary methods are powerful tools to capture boundary condition effects with Cartesian grids. They are based on modifying the discretization of governing equations for the grids near the interface to mimic the effect of the boundary conditions. 
 Although few IBM formulations exist that involve moving interfaces, they do not have a general framework to capture the evolution of rigid boundaries as controlled by reactions (instead they may have a prescribed interfacial velocity). On the other hand, existing methods that have used  LS   for reactive transport problems have not taken advantage of an  immersed boundary formulation to implement high order accuracy boundary conditions implicitly. For example, \cite{li2010three} used a first order interpolation to implement boundary conditions which is explicit in nature. 
 
 In this work we propose a framework that couples IBM and LS to model reactive transport in complex topologies with moving interfaces on a Cartesian grid; we also propose a novel propagation formula for the interface velocity. The IBM formulation is based on the high-order method developed by Yousefzadeh and Battiato~\cite{yousefzadeh2019high, yousefzadeh2020numerical} that is capable of solving Dirichlet, Neumann and Robin boundary conditions. The interface velocity propagation scheme of Li \emph{et al.}~\cite{li2008level} is improved and extended to ensure a consistent minimum second order of accuracy in space throughout the numerical framework.
 
 The paper is organized as follows.  In Section \ref{Sec:Problem_form}, we present the governing equations. The details of the numerical implementation of the proposed algorithm are presented in Section \ref{sec:numeric_method}. In Section \ref{sec:results}, we apply the proposed method to model, first, the dissolution of a  calcite grain exposed to acid injection and, second, the diffusion-limited dendritic growth of a solid nucleus in an over-saturated solute. We conclude in Section \ref{sec:conclusions} with a summary of our work and main results.

\section{Problem Formulation}\label{Sec:Problem_form}

The conservation of momentum of a Newtonian, viscous, incompressible fluid with constant properties flowing around an impermeable solid obstacle is governed by the dimensionless Navier-Stokes equations
\begin{align}\label{eq:NavierStokes}
\dfrac{\partial u_i}{\partial t}+  \dfrac{\partial u_i u_j}{\partial x_j}  =-\dfrac{\partial p}{\partial x_i} +\dfrac{1}{\mathit{Re}}\dfrac{\partial}{\partial x_j} \left(\dfrac{\partial u_i}{\partial x_j}\right) ,   \qquad  \dfrac{\partial u_i}{\partial x_i}=0, 
\end{align}
where  $p$ [-] and $u_i$ [-] are the dimensionless  fluid pressure and velocity component in the coordinate direction $i$. Also, 
\begin{align}\label{eq:reynolds}
Re = \dfrac{UL}{\nu}
\end{align}
is the Reynolds number, where $U$ [LT$^{-1}$], $L$ [L] and $\nu$ [L$^{2}$T$^{-1}$] are a characteristic velocity, a characteristic length and the fluid kinematic viscosity. 
Equations \eqref{eq:NavierStokes} are subject to the no-slip boundary condition at the fluid-solid interface, i.e. the immersed boundary, $\Gamma_{\tiny{\mbox{IB}}}$, 
\begin{align}\label{bc:stokes}
 \mathbf u =\textbf 0, \quad \mathbf x \in \Gamma_{\tiny{\mbox{IB}}}.
\end{align}

The transport of a  species  undergoing a heterogeneous reaction (dissolution or precipitation) on the boundary of an immersed solid object can be described by an  advection-diffusion equation for the dimensionless  species concentration $C$ [-]
\begin{align}\label{eq:transport}
\dfrac{\partial C}{\partial t}+ \dfrac{\partial (u_j C) }{\partial x_j}=\dfrac{1}{Pe}\dfrac{\partial^2 C}{\partial x_j^2},
\end{align}
where $C = \hat{C}/C_c$, $\hat{C}$ $[ML^{-3}]$ and $C_c$ $[ML^{-3}]$ are the species concentration and its characteristic value, respectively, and $Pe$ is the P\'{e}clet number and is defined as
\begin{align}\label{eq:peclet}
Pe = \dfrac{UL}{D},
\end{align}
with $D$ [L$^2$T$^{-1}$] the molecular diffusion coefficient. Equation \eqref{eq:transport} is subject to  reactive boundary conditions on the  solid-fluid interface, $\Gamma_{\tiny{\mbox{IB}}}$
\begin{align}\label{bc:reactiveBC}
- n_i \dfrac{\partial C}{\partial x_i} =Da C_{\Gamma}, \quad \mathbf x\in \Gamma_{\tiny{\mbox{IB}}},
\end{align}
where $C_{\Gamma}$ [-] is the dimensionless interface concentration, $Da$ [-] and $n_i$ are the Damk\"{o}hler number and the component of the normal $\mathbf n$ to the solid-fluid interface in the $i$ direction or towards the fluid region. The Damk\"{o}hler number is defined as
\begin{align}
Da = \dfrac{KL}{D},
\end{align}
where $K$ $[LT^{-1}]$ is the reaction rate constant. The speed of the moving solid boundary due to precipitation/dissolution reaction is given by \cite{molins2017mineralogical, li2008level}, 
\begin{align}\label{eq:IntvelocityDim}
\hat{u}_{\Gamma} = -K\hat{C}_{\Gamma}V_m, \quad  \text{at}\;\Gamma_{\tiny{\mbox{IB}}},
\end{align}
where $V_m$ $[M^{-1}L^3]$ is the species molar volume, $\hat{C}_{\Gamma}$ [$ML^{-3}$] is the interface concentration and $\mathbf{\hat{u}_\Gamma}=\hat{u}_\Gamma \mathbf n$ [$LT^{-1}$] is the interface velocity vector. Equation \eqref{eq:IntvelocityDim} can be expressed in  dimensionless form as
\begin{align}\label{eq:IntvelocityDimless}
\mathbf{u}_{\Gamma} = -\dfrac{Da}{Pe}V_m C_c  C_{\Gamma}\mathbf{n}, \quad  \text{at}\;\Gamma_{\tiny{\mbox{IB}}}.
\end{align}
Without loss of generality, the dimensionless quantity $V_m C_c$  [$M^0L^0$] is set to one in all simulations.
\subsection{Level Set Representation of Evolving Reactive Interface}
Accurately formulating and tracking the motion of an interface in reactive transport modeling at the pore-scale  is  critical to capture mass fluxes across a reactive and moving interface. The Level-set method \cite{osher1988fronts} is a powerful approach to attain this objective, although mass losses associated with the nonconservative nature of  LS formulation require a very fine  grid near the interface. Recently, a number of studies have developed methods to capture  local mass conservation in LS methods~\cite{Jettestuen2021-wy, Ge2018-eq}. The ghost cell immersed boundary method (GCIBM) treatment of the boundary conditions preserves the discontinuous nature of the variables across the solid-fluid interface in a sharp manner. Such features are critical in applications where the mass, heat or momentum boundary layer is thin and comparable to the support thickness of the smoothed delta function used in diffuse interface methods. 

The level-set method describes the interface, $\Gamma_{\tiny{\mbox{IB}}}$, as the zero-level contour of a higher dimensional function $\phi$. For practical reasons, the level-set function is defined as the signed distance function  from the interface $\Gamma_{\tiny{\mbox{IB}}}$. Specifically, $\phi$ is equal to the minimum distance between each point in the domain and the interface, where such a distance is negative or positive if the points belong to the solid or fluid subdomains, respectively,

\begin{equation}\label{eq:LSdefine}
\phi(\mathbf x) =
    \left\{
    	\begin{array}{ll}
    		\delta {(\mathbf x)}  & \mathbf x\in \Omega_{\tiny{\mbox{fluid}}} \\
    		-\delta{(\mathbf x)} & \mathbf x\in \Omega_{\tiny{\mbox{solid}}}  \\
    		0 & \mathbf x\in \Gamma_{\tiny{\mbox{IB}}}
    	\end{array}
    \right.
\end{equation}
where $\delta$ is the minimum (Euclidean) distance between $\mathbf x$ and the interface. The level-set function is also used to compute the normal to the interface as
\begin{align}\label{eq:normal}
\mathbf n = \dfrac{\nabla \phi}{|\nabla \phi|}.
\end{align}
The interface velocity, defined by \eqref{eq:IntvelocityDimless} and due to the chemical reaction at the solid-fluid interface \eqref{bc:reactiveBC}, satisfies the level-set equation
\begin{align}\label{eq:levelset}
\dfrac{\partial \phi}{\partial t}+ \mathbf{u}_{\Gamma}\cdot\nabla \phi=0, 
\end{align}
where  $\mathbf{u}_{\Gamma}$ is the dimensionless interface velocity.
It is worth emphasizing that the computation of the normal vector \eqref{eq:normal}, the interface velocity \eqref{eq:IntvelocityDimless} and reactive boundary condition \eqref{bc:reactiveBC} at the interface all depend upon the level set function property to remain a signed distance function. However, in general,  the level set function ceases to hold its signed distance function property as it evolves in time  \cite{osher2001level}. A reinitialization process is therefore adopted at every level-set time step to project the new level set to a signed distance function, while preserving the zero level set location according to \cite{sussman1994level}
\begin{align}\label{eq:reinit}
\dfrac{\partial \phi}{\partial \tau} + \text{sgn}(\phi^0)(1-|\nabla \phi|)=0,
\end{align}
where $\tau$ is a pseudo-time step and $\phi^0$ is the level set function obtained after solving \eqref{eq:levelset}. Equation \eqref{eq:reinit} needs to be solved up to the pseudo-time steady state.  
\section{Numerical Methodology}\label{sec:numeric_method}
\subsection{Finite Volume Discretization}\label{sec:Finite_vol}
Equations \eqref{eq:NavierStokes}-\eqref{eq:transport} are discretized using a finite-volume scheme \cite{ferziger2012computational,versteeg2007introduction} on a structured Cartesian grid. {An Euler backward time integration scheme is used for temporal discretization.}

The NS equations \eqref{eq:NavierStokes} are solved in the primitive variables ($\mathbf u, p$) formulation. The primitive variables are defined on a staggered grid. This avoids the checkerboard patterns in the solution \cite{harlow1965numerical}. In staggered grids, the velocities are stored at the grid faces and the scalar variables (e.g. pressure, temperature, concentration) are stored at the cell centers. Hence, a $s+1$ Cartesian grid set is required in a $s$-dimensional problem. The pressure-implicit with splitting of operators (PISO) algorithm, a non-iterative  method to solve the unsteady Navier-Stokes equation with large time steps, is adopted for pressure-velocity coupling \cite{issa1986solution} and satisfies mass conservation using predictor-corrector steps. The convection and viscous terms in \eqref{eq:NavierStokes} are discretized using third-order quadratic upstream interpolation for convective kinematics (QUICK) \cite{leonard1979stable,hayase1992consistently} and a second-order central difference schemes, respectively. An identical spatial discretization is adopted to discretize the convection and diffusion terms in the scalar transport equation \eqref{eq:transport}.
Hamilton-Jacobi weighted essentially non-oscillator (WENO) scheme \cite{liu1994weighted, fedkiw1999non} is used to discretize both the level set and the reinitilization equations \eqref{eq:levelset}-\eqref{eq:reinit} in space. Their temporal discretization is performed by either total variation diminishing Runge-Kutta third-order (TVD-RK3) or total variation diminishing Runge-Kutta fourth-order (TVD-RK4) \cite{shu1988efficient}.  A comprehensive review of  strategies to solve the level set and reinitialization equations is given in \cite{gibou2018review, osher2004level}. \noah{Details on the temporal discretization can be found in~\ref{sec:press-vel-coupling} and~\ref{app:other-temp}.}
%


\subsection{Immersed Boundary Treatment}\label{sec:IBM_sec}
The level set tracks the evolution of the interface, as well as the information about the normal to the interface. In order to implement  equations \eqref{eq:NavierStokes} and \eqref{eq:transport} on Cartesian grids, their associated boundary conditions need to be properly treated. In this study, we use a variation of the GCIBM developed in \cite{yousefzadeh2019high, yousefzadeh2020numerical} to implement boundary conditions.  

The method is based on a pre-classification of the type of cell: fluid, solid or ghost. Fluid cells are the cells whose centers are located inside the flow domain; solid cells have their centers inside the solids and are not directly adjacent to a fluid cell; ghost cells are the cells inside the solid with at least one neighboring fluid cell. A flag variable is constructed to assign each cell to a specific class (liquid, solid, ghost).   The values of state variables inside solids, fluid  and ghost cells are treated differently.

Instead of using more classical approaches, e.g. the ray-casting algorithm \cite{yousefzadeh2019high}, we note that the Level Set function itself provides a natural framework for cell classification. In particular, if the cell center is in the flow domain then it will have a negative level set value ($\phi<0$). In solid cells,   $\phi> 0$.  If the level set has a zero value ($\phi = 0$), the cell center is on the boundary and will be flagged as fluid cell. Since the solid interface perfectly aligns with the cell center, approximation or interpolation is not required. The ghost cells are identified as the solid cells with at least one fluid cell neighbor. The detailed steps of the identification process are outlined in  Algorithm \ref{alg:PointID}.
\begin{algorithm}
\caption{Point Identification Algorithm}\label{alg:PointID}
\begin{algorithmic}[1]
\Function{PointID}{$X,Y,\phi$}\Comment{The type of each grid point}
\ForAll{$(x_i,y_j) \in \{(X,Y)\}$}
	\State $\phi(x_i,y_j)\gets$ level set value at $(x_i,y_j)$
	\If {$\phi(x_i,y_j) \ge 0$}
    	\State $F(i,j) \gets 1$  \Comment{flag value $1$ is for fluid points}
        \State $(x_i,y_j)\in {(X_f,Y_f)}$
     \ElsIf {$\phi(x_i,y_j)<0$}  
         \State $F(i,j) \gets -1$  \Comment{flag value $-1$ is for solid points}
         \State $(x_i,y_j)\in {(X_s,Y_s)}$
    \EndIf
\EndFor\label{PointIDFor}
\ForAll{$(x_i,y_j)\in {(X_s,Y_s)}$}
	\If {$F(i+1,j)+F(i-1,j)+F(i,j+1)+F(i,j-1) \neq -4$} 
    	\State $F(i,j) \gets 0$  \Comment{flag value $0$ is for ghost points}
         \State $(x_i,y_j)\in {(X_g,Y_g)}$
    \EndIf
\EndFor
\State \textbf{return} $F$
\EndFunction
\end{algorithmic}
\end{algorithm}
The tagging process will be performed at every time step to reflect the changes in solid-liquid interface. Once the tagging process is completed (see Figure~\ref{fig-tagging}), an appropriate formulation for each class of the cells will be used to update the state variables. The algebraic equation for each variable in the fluid cells will be obtained by discretizing the governing PDEs. The values of all state variables in the solid cells are set to be constant, however, one may prefer to exclude the solid cells from the computations as they are completely decoupled from the rest of the domain. Finally, the algebraic equation for the ghost cell values are constructed in such a way that their coupling with the fluid cells will result in correct boundary condition implementation. Details of the implementation can be found in \cite{yousefzadeh2019high, yousefzadeh2020numerical}. Specifically, the GCIBM developed in \cite{yousefzadeh2019high} is capable of treating the general boundary condition of the form 
\begin{align}\label{bc:general2}
&- \mathbf{n}\cdot \alpha\dfrac{\partial \psi}{\partial r_n} =\beta\psi+ q, \quad  r_n=0,
\end{align}
for a generic state variable $\psi$. Both a Dirichlet boundary condition in the NS Equation \eqref{bc:stokes} and a reactive boundary condition in the transport equation \eqref{bc:reactiveBC} can be modeled through \eqref{bc:general2} by defining the proper values for $\alpha$, $\beta$ and $q$.

For a more detailed discussion on the implementation of GCIBM, we refer the interested reader to \cite{yousefzadeh2019high, yousefzadeh2020numerical}. \noah{For dissolution/precipitation cases, as the interface moves and shrinks, solid or ghost cells at the prevailing time step might become ``new" (swept) cells at the new time step. In the current GCIBM algorithm, no special treatment has been used to approximate the values in the swept (fresh/dead) cells: the velocity value in the swept cell is set to zero at the new time step.  As widely reported in the literature, the generation of fresh and dead cells due to moving or deforming bodies will violate local mass conservation. Seo \& Mittal~\cite{Seo2011-vq} derived the volume conservation error $err_{\Delta V}$ for each individual fresh or dead cell as}

\noah{
    \begin{equation}
       err_{\Delta V} = \frac{\Delta V}{\Delta t}  \left\vert  1 -CFL_b \right\vert, \label{eq:err_delV}
    \end{equation}}
\noah{where $\Delta V = \Delta {x}^2$ or $\Delta {x}^3$ for 2-D and 3-D cases, respectively, and $CFL_b$ is the local Courant number. Based on equation~\eqref{eq:err_delV}, $err_{\Delta V}$ is proportional to $\Delta V$ and inversely proportional to $\Delta t$. Assuming $\left\vert  1 - CFL_b \right\vert \sim 1$, $ \frac{\Delta V}{\Delta t} < 1$ is required to reduce the errors due to the violation of mass conservation during the generation of fresh and dead cells. In this work, we ensure that $\frac{\Delta V}{\Delta t} < 1$ is satisfied for all our simulations. As such, the error due to local mass conservation violations is insignificant. In addition to controlling $\frac{\Delta V}{\Delta t}$, other methods, such as cut-cell-based discretization [54] and interpolation methods [55], have been proposed to enforce local mass conservation without constraining $\frac{\Delta V}{\Delta t}$.}


\subsection{Level Set-Immersed Boundary Method}\label{sec:LSIBM}
A level set function is used to implicitly capture the interface separating the fluid and solid ($\Gamma_{\tiny{\mbox{IB}}}$). The evolution of the interface (boundary) is governed by equation \eqref{eq:levelset}, where equation \eqref{eq:IntvelocityDimless} defines  the velocity of the interface in the normal direction to  $\Gamma_{\tiny{\mbox{IB}}}$. 

Once the interface velocity, $u_n$ is found, the level set function is advanced in time from $t^n$ to $t^{n+1}$ by solving equation \eqref{eq:levelset}. Subsequently, the reinitialization equation \eqref{eq:reinit} is solved to ensure that the updated level set remains a signed distance function. The new level set defines the current location of the interface at time $t^{n+1}$, and can be used to obtain the normal vector to the interface, $\mathbf n$. The flow and transport equations are finally solved to update  velocity, pressure and concentrations using the GCIBM. It is worth emphasizing that Eqs. \eqref{eq:levelset} and \eqref{eq:reinit} need to be solved only within a narrow region in proximity of the interface, $|\phi| \leq \varepsilon$, where $\varepsilon$ is larger than the stencil required for WENO scheme, i.e. $\varepsilon \geq 5\Delta x$. No special treatment is required for $\abs{\phi} = \varepsilon$.

\subsubsection{Interface Velocity Propagation}\label{sec:LSvel}

\begin{figure}
    \centering

    \begin{subfigure}[ht]{0.45\textwidth}
      \caption{}
      \begin{center}
        \includegraphics[clip, trim={3cm 6cm 2cm 5.0cm},width=\textwidth]{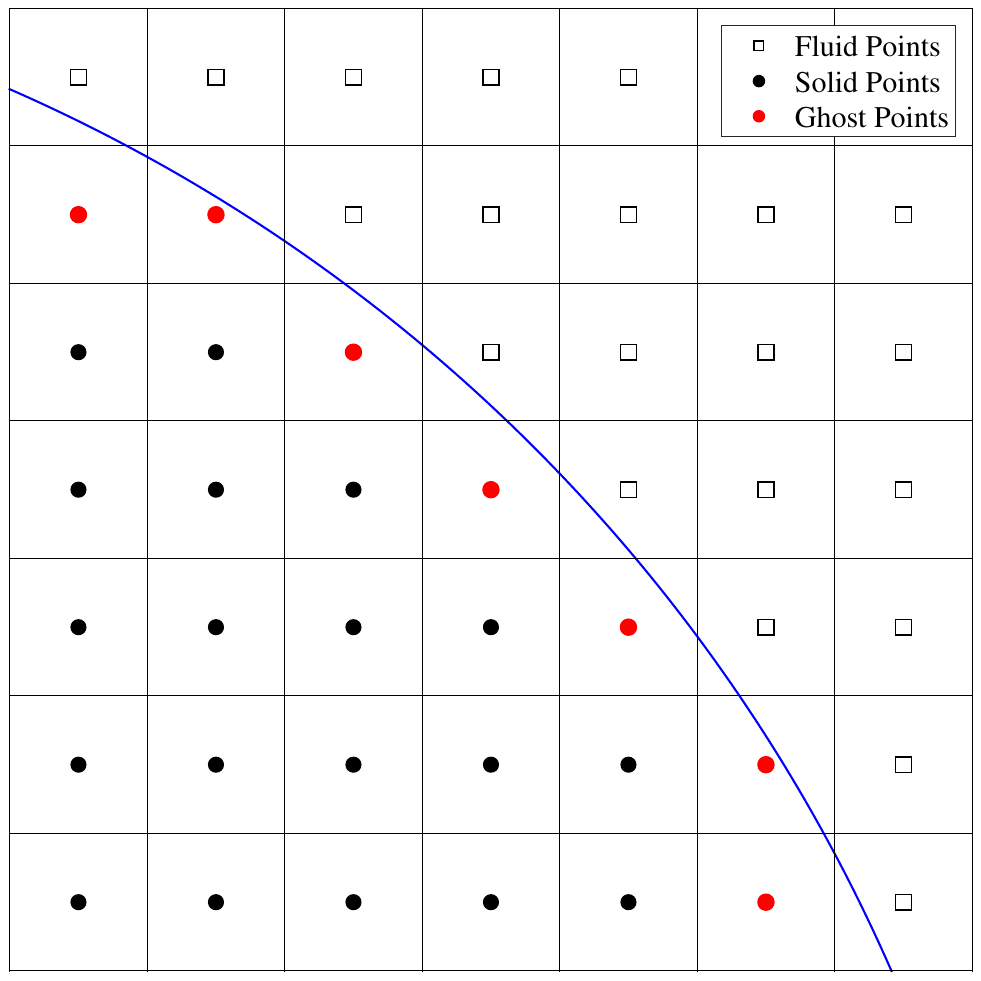}
      \end{center}
      \label{fig-tagging}
    \end{subfigure}
    \hfill
    \begin{subfigure}[ht]{0.45\textwidth}
    \caption{}
      \begin{center}
        \includegraphics[width=\textwidth]{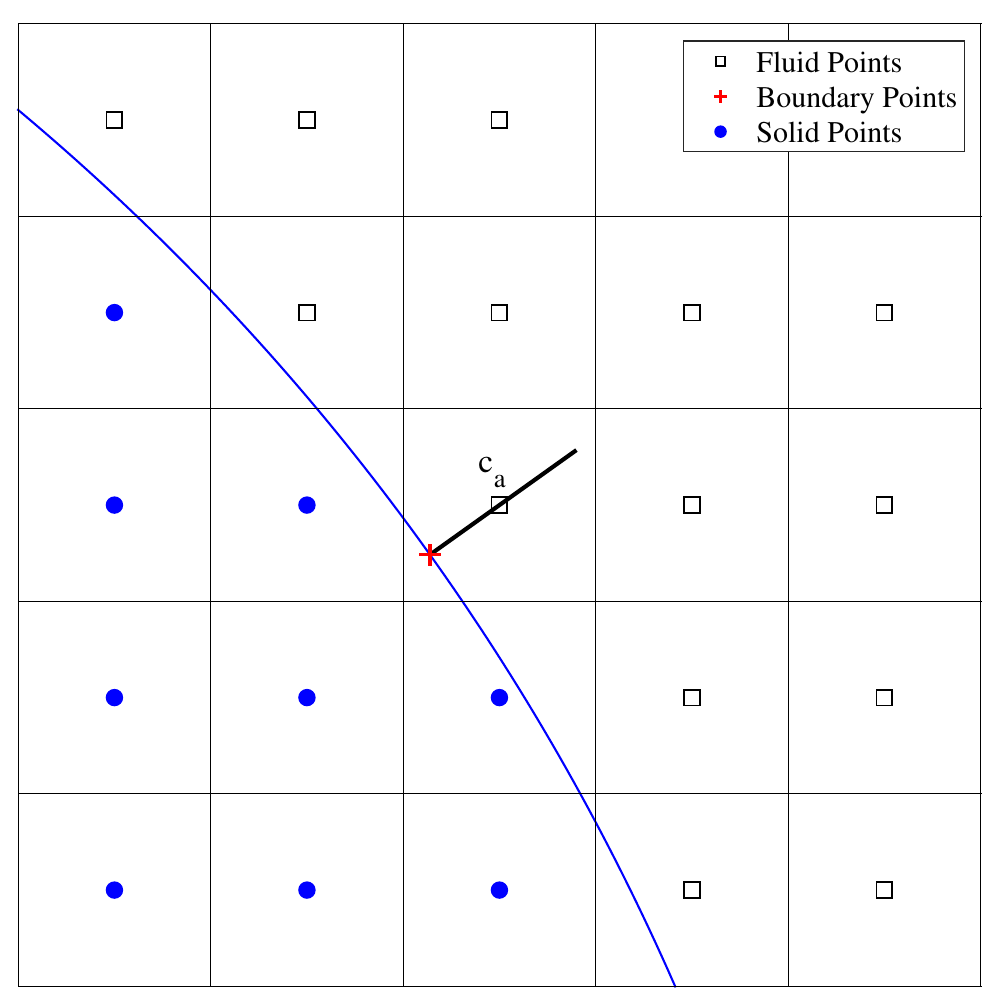}
      \end{center}
      \label{LiFormula}
    \end{subfigure}
     
     \begin{subfigure}[ht]{0.45\textwidth}
      \caption{}
      \begin{center}
        \includegraphics[width=\textwidth]{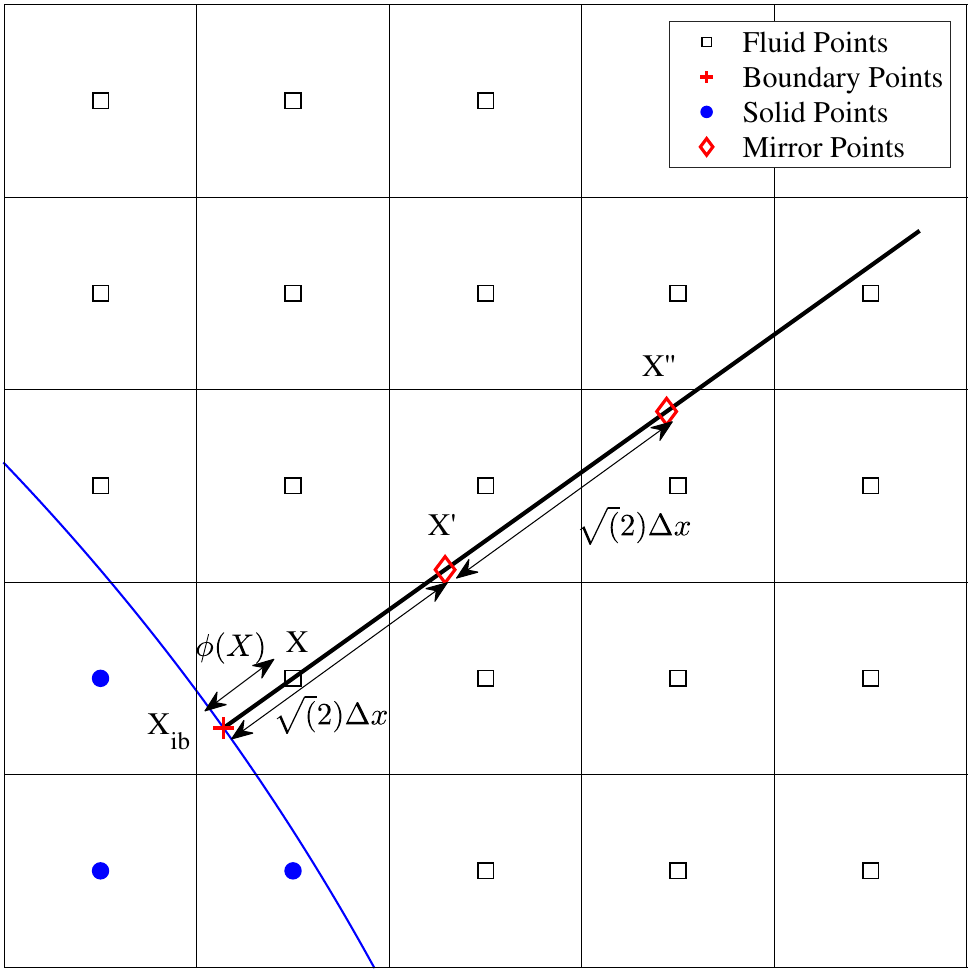}
      \end{center}
      \label{OurFormula}
     \end{subfigure}
     \hfill
     \begin{subfigure}[ht]{0.45\textwidth}
      \caption{}
      \begin{center}
        \includegraphics[width=\textwidth]{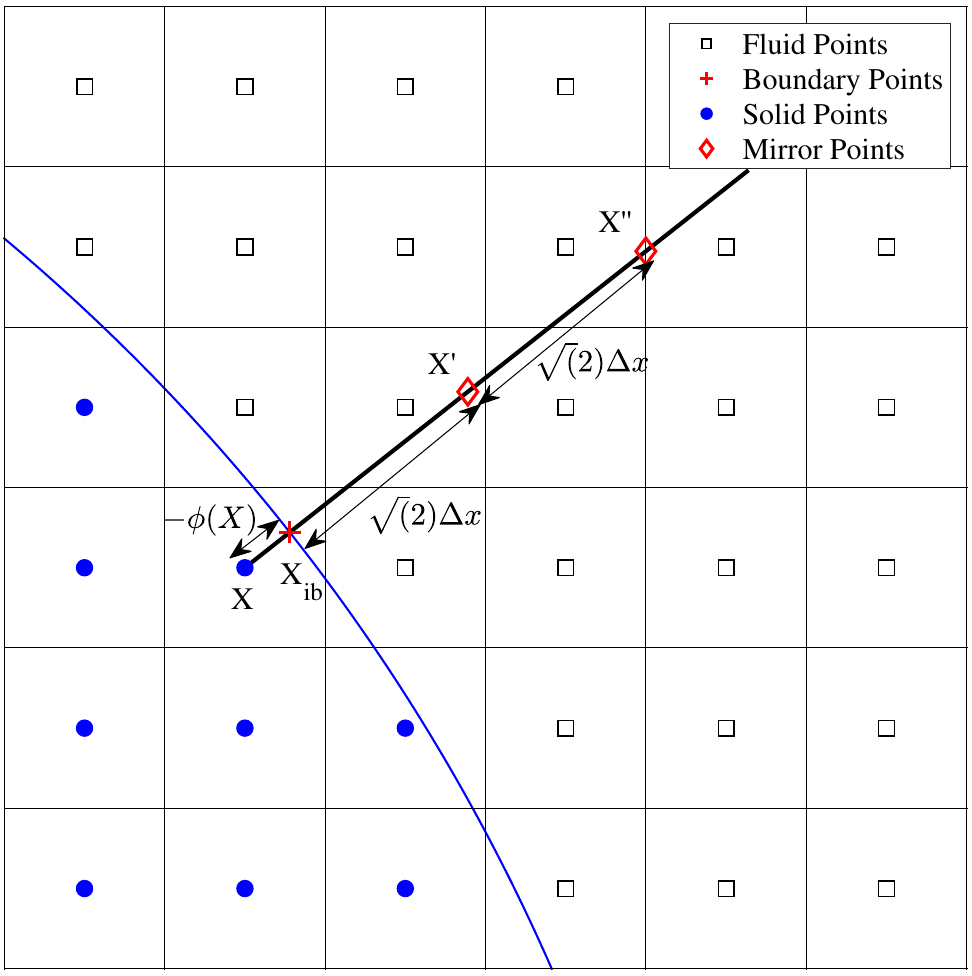}
      \end{center}
      
      \label{OurFormula2}
     \end{subfigure}
     \caption{(I) A 2D schematic representation of different type of the points in the domain. (II) The classification of points and stencil used in Li et al. \cite{li2008level} and the stencil used in velocity extrapolation for (III) fluid points and (IV) solid points}
     \label{OurFormulas}
\end{figure}

Although Equation~\eqref{eq:levelset} assumes the velocity field $u_n$ is defined in the entire computational domain, the velocity is only properly defined at the interface. This calls for propagating (extrapolating) the velocity to the narrowband around the interface ($\phi = 0$). A common approach to propagate the interface velocity is the method introduced by \cite{fedkiw1999non}. Fedkiw \emph{et al.} \cite{fedkiw1999non} write 
\begin{align}\label{eq:ghostfluid}
\dfrac{\partial \lambda}{\partial \tau} \pm n_i\dfrac{\partial \lambda }{\partial x_i}=0
\end{align}
to propagate a generic variable $\lambda$, where the plus and minus signs are used when  $\phi>0$ and  $\phi<0$, respectively.  Aslam  \cite{aslam2004partial} extended this method   by introducing the concept of PDE extrapolation, where a higher order PDE is solved to perform extrapolation with higher order of accuracy.  Li \emph{et al.} \cite{li2008level, li2010three} employed the ghost fluid method to propagate the interface velocity in a narrowband of the interface inside  the solid domain. In the fluid domain, instead of solving the extrapolation equation \eqref{eq:ghostfluid}, they derive an explicit expression for the interface velocity field at point $a$ (see Fig. \ref{LiFormula}) as
\begin{align}\label{eq:veloLi}
u^a_n = \dfrac{Da}{Pe}\left(\dfrac{1}{1+\phi_a Da}\right)C_a.
\end{align}
The velocity field in the fluid domain serves as the boundary condition for equation \eqref{eq:ghostfluid}. Equation \eqref{eq:veloLi} is derived based on the fact that  equation \eqref{eq:ghostfluid} is solved at  steady state, (i.e. $n_i\partial \lambda/\partial x_i=0$). They used a first order approximation of the derivative to perform the constant extrapolation as shown in  Figure \ref{LiFormula}. 

However, the interface velocity propagation algorithm by \cite{li2008level} employs two separate approaches for the solid and fluid sub-domains. Moreover, Equation \eqref{eq:veloLi} is  first-order accurate. The formula does not suggest a local formulation for the points farther from the interface. This becomes critical specially for the cases where there is a high concentration gradient near the interface.
Here, we improve upon the formulation of  \cite{li2008level, li2010three} by (i) deriving a new expression for interface velocity  to include both the fluid and solid domain and (ii) employing a different stencil in the extrapolation to obtain a higher order of accuracy while ensuring a more local formulation. 

The steady state solution of  equation \eqref{eq:ghostfluid} for the interface velocity can be expressed as
\begin{align}\label{eq:SSvel}
n_i\dfrac{\partial u_n }{\partial x_i}=0. 
\end{align}
Equation \eqref{eq:SSvel} indicates that the normal velocity should be constant along the normal vectors to the interface. Therefore, for any grid point $X = (x,y)$ in the neighborhood of the interface ($|\phi|\le \varepsilon$), we need to find the corresponding boundary points (i.e. the intersection of the interface and the line passing through $X$ with its slope equal to the normal vector $n_X$). The boundary points can be obtained as follows
\begin{align}
X_{\tiny{\mbox{IB}}} = (x_{\tiny{\mbox{IB}}}, y_{\tiny{\mbox{IB}}}) = (x - n_x \phi_{(x,y)}, y - n_y \phi_{(x,y)}).
\end{align}
The level set velocity at $X = (x,y)$ is equal to the velocity at its corresponding $X_{\tiny{\mbox{IB}}} = (x_{\tiny{\mbox{IB}}}, y_{\tiny{\mbox{IB}}})$, 
\begin{align}\label{eq:IntvelocityIB}
\mathbf{u}^{\Gamma}_{(x,y)} = \mathbf{u}_{\tiny{\mbox{IB}}} = -\dfrac{Da}{Pe}V_m C_c C_{\tiny{\mbox{IB}}} \cdot \mathbf{n}.
\end{align}
The concentration at the interface, $C_{\tiny{\mbox{IB}}}$, is needed to fully describe the velocity at $X$ and it is reconstructed using two mirror points in the fluid domain and a one-sided second-order finite difference expression for the concentration derivative in the normal direction. A second-order approximation of the derivative requires the values at three points. Equation \eqref{bc:reactiveBC} is employed to close the system. If the mirror point does not coincide with a grid point, its value is obtained by a bilinear interpolation. The two mirror points are defined as
\begin{align}\label{eq:LSIBpoints}
&X' = (x', y') = (x_{\tiny{\mbox{IB}}} + \ell n_x ,y_{\tiny{\mbox{IB}}} + \ell n_y)  \nonumber \\
&X'' = (x'', y'') = (x' + \ell n_x ,y' + \ell n_y)
\end{align}
where $\ell  = \sqrt{2}\Delta x$ ensures that the mirror point is surrounded by four points in the fluid domain as shown in Figures \ref{OurFormula} and \ref{OurFormula2}. 

The concentration value at the interface can be obtained from
\begin{align}\label{eq:IntConcent}
C_{\tiny{\mbox{IB}}} = \dfrac{4C_{X'}-C_{X''}}{2\ell Da  + 3}.
\end{align}
The derivation details can be found in~\ref{sec:2ndorder_CIB}. Finally the level set velocity vector can be calculated as
\begin{align}\label{eq:Intvelfinal}
\mathbf{u}^{\Gamma}_{(x,y)} = -\dfrac{Da}{Pe}V_mC_c\left (\dfrac{4C_{X'}-C_{X''}}{2\ell Da + 3}\right )(n_x,n_y) .
\end{align}
The current formulation provides a general expression for both solid and fluid points. 

\subsection{Discussion on the Practical Aspects of LS-IBM}\label{sec:Timestep}
The IBM flow and transport solver is based on an implicit method. As a result, there is no severe restriction on the time step. Yet, a reasonably small time step, close to the viscous (diffusive) and Courant–Friedrichs–Lewy (CFL) (advective) time steps, is recommended for faster convergence. This choice also honors  the physical time scales.  

The GCIBM requires that the grid Damk\"{o}hler number ($Da_{\Delta x} = K\Delta x/D$) be less than 2 due to the explicit dervivation of coefficients for polynomial approximation of the boundary conditions. In particular, in this method,  the state variable of the Robin boundary condition is approximated with a polynomial function. The coefficients of the function are derived explicitly with a constant factor of $(2-Da_{\Delta x})^{-1}$ where $Da_{\Delta x}= K\Delta x/D$ refers to numerical Damk\"{o}hler number. Since {$(2-Da_{\Delta x}) > 0$} to prevent an ill-conditioned matrix,  then {$Da_{\Delta x} < 2$}. This poses restrictions on the grid size close to the interface. Table \ref{tab:timesteps} summarizes general time step and grid size restrictions, where the safety factor $\lambda_{cfl}$ for ensuring stability in Table \ref{tab:timesteps}  is between 0 and 1. \noah{We emphasize that some of the listed grid size and time step restrictions may be attenuated by using implicit schemes. For example, since in the current implementation diffusive terms are treated implicitly, the viscous time-step constraint is eliminated.}

\begin{table}[htbp]
\begin{center}
\caption {\noah{Possible grid size and time step restrictions. Viscous, CFL and Diffusion time-step constraints are attenuated with implicit treatment.}} \label{tab:timesteps} 
\begin{tabular}{l*{6}{c}r}
\hline \\          
Viscous time step       &$\Delta t \leq \Delta t_{\nu} = \dfrac{Re}{\dfrac{1}{\Delta x^2}+\dfrac{1}{\Delta y^2}}$   \\\\
CFL time step					& $\Delta t \leq \Delta t_{cfl} = \dfrac{\Delta x}{u_{max}}$\\\\
Diffusion time step       &$\Delta t \leq \Delta t_D = \dfrac{Pe}{\dfrac{1}{\Delta x^2}+\dfrac{1}{\Delta y^2}}$   \\\\
IBM grid size       &$\Delta x \leq \Delta x_{IBM} = \dfrac{2K}{D}$   \\\\
LS time step       &$\Delta t \leq \Delta t_{LS} = \dfrac{\lambda_{cfl}\Delta x_{min}Pe}{Da}$   \\\\

\hline
\end{tabular}
\end{center}
\end{table}

Also, since the movement of the interface due to reaction is much smaller than flow and transport processes, one may not solve the level set equation (i.e. update the boundary geometry) at every flow and transport time step, i.e. the time step in \eqref{eq:levelset} is a multiple of the time step in  \eqref{eq:NavierStokes} and \eqref{eq:transport}, namely, $\Delta t_{LS} = m \Delta t$, with $m$ an integer.
\section{Numerical Implementation}\label{sec:results}
%
In the following, we validate the two-dimensional implementation of
the scheme against two benchmark problems: a 2D cylindrical calcite grain dissolution problem \cite{molins2020simulation} and a precipitation-driven dendrite growth problem.
\subsection{Dissolution of a Cylindrical Grain}\label{sec:results2}
\begin{figure}[H]
  \begin{center}
    \includegraphics[trim={2.5cm 5.7cm 1cm 3cm},clip,width=6in]{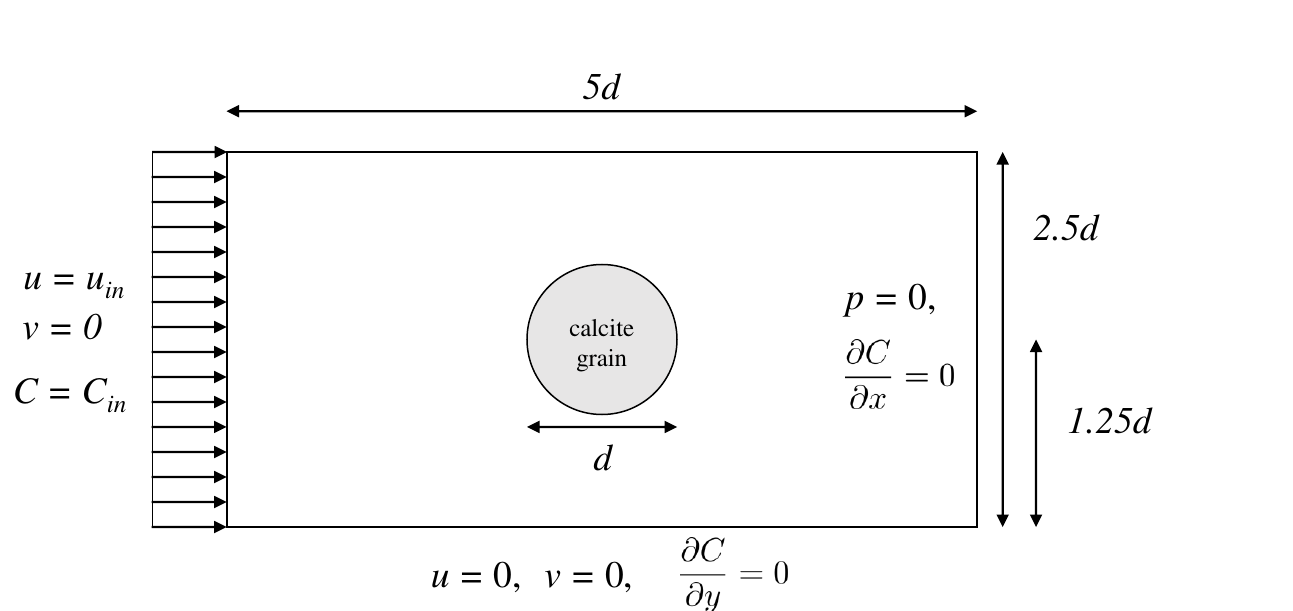}
  \end{center}
  \caption{Computational domain and BCs for reacting grain}
  \label{fig:schemDomain}
\end{figure}

We consider  flow and transport around a 2D cylindrical calcite grain, fully immersed in an acidic fluid. The reaction at the solid-fluid interface dissolves the grain. We consider the benchmark problem  setup of Molins \emph{et al.} \cite{molins2020simulation}. The total length and width of the computational domain is $5d$ and $2.5d$ respectively, where $d$ is the initial diameter of the grain. The grain is placed at the center of the domain as shown in Figure \ref{fig:schemDomain}. The simulation parameters are summarized in  Table \ref{tab:simParam}.

\begin{table}[H]
\begin{center}
\caption {Simulation parameters for dissolving grain} \label{tab:simParam} 
\begin{tabular}{l*{9}{c}r}
\hline           
Reynolds number       & {$Re = 0.6$}   \\
Inlet velocity        &$u_{in} = 1$   \\
Inlet concentration   &$C_{in} = 1$   \\
Grid size	          &$\Delta x = 0.005$\\
Time step             &{$\Delta t = 10^{-5}$~--~$10^{-3}$}   \\
Grain diameter        &$d = 0.4$   \\
\hline
\end{tabular}
\end{center}
\end{table}

In order to test the robustness and capabilities of the LSIBM we performed the dissolution of the solid grain at multiple $Pe$ and $Da$. Table \ref{tab:Da-Pe} summarizes the dimensionless groups used in the simulations to analyze both convergence, accuracy and stability of the solver, as well as to validate the solver against benchmark cases.

\begin{table}[H]
\begin{center}
\caption {$(Da, Pe)$ pairs} \label{tab:Da-Pe} 
\begin{tabular}{l*{6}{c}r}
\hline
Case & $Da$ & $Pe$ & $Da_I = Da Pe^{-1}$   \\
\hline
(I) & $0.178$ & $6$ & $0.03$ \\
(II) & $17.8$ & $0.6$ & $29.67$ \\
(III) & $1.78$ & $6$ & $0.3$ \\
(IV) & $1780$ & $6$ & $296.67$ \\
(V) & $178$ & $600$ & $0.3$ \\
(VI) & $17800$ & $600$ & $29.67$ \\
\hline
\end{tabular}
\end{center}
\end{table}

\subsubsection{Grid convergence, accuracy and stability analysis}
\begin{figure}[H]
     \centering
     \begin{subfigure}[ht]{0.48\textwidth}
     \caption{}
      \begin{center}
        \includegraphics[width=\textwidth]{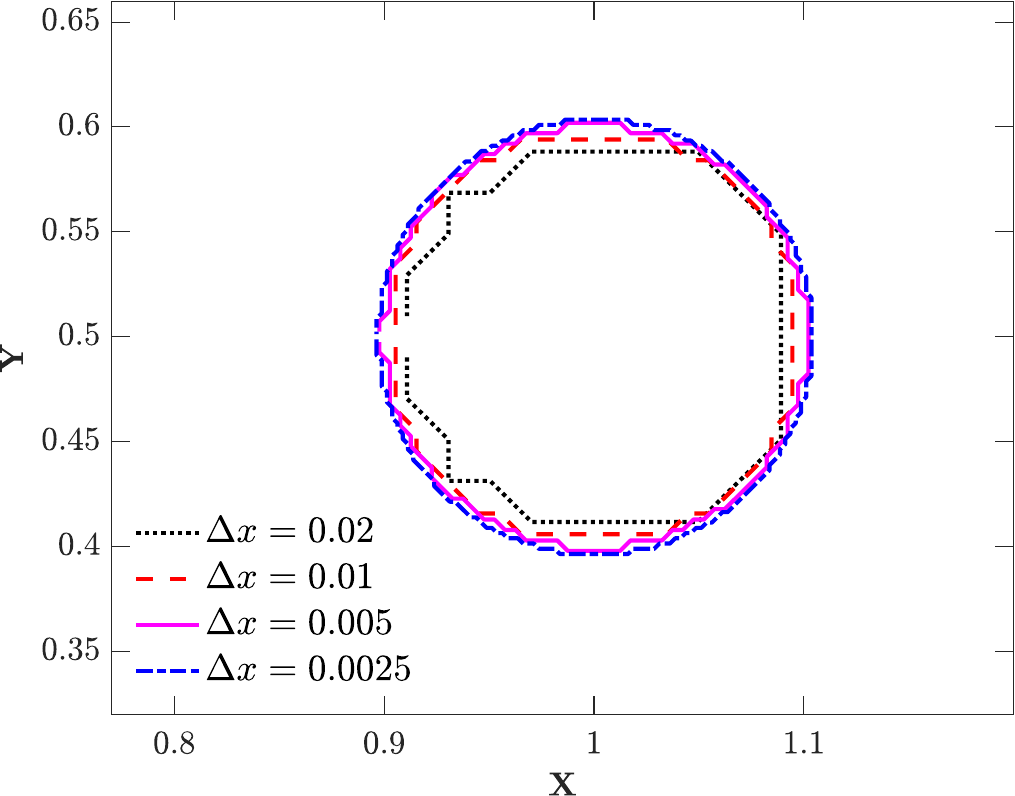}
      \end{center}
      \label{fig:err-case1}
     \end{subfigure}
     \hfill
     \begin{subfigure}[ht]{0.48\textwidth}
     \caption{}
      \begin{center}
        \includegraphics[width=\textwidth]{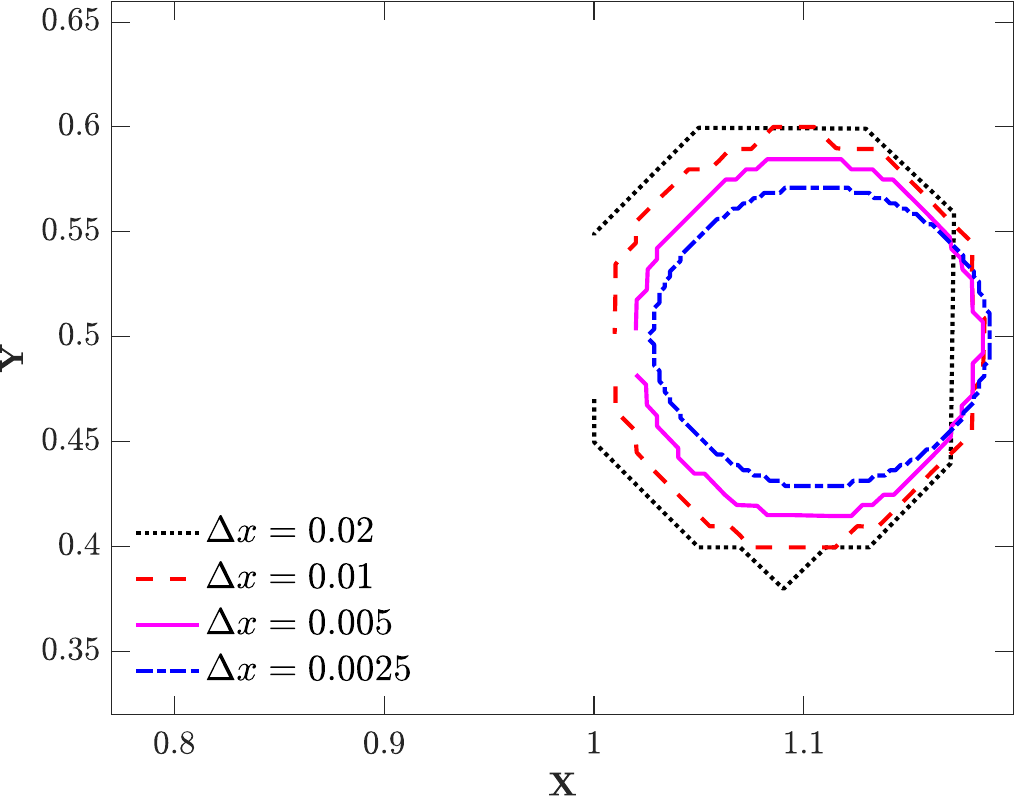}
      \end{center}
      \label{fig:err-case2}
     \end{subfigure}

     \begin{subfigure}[ht]{0.48\textwidth}
      \caption{}
      \begin{center}
        \includegraphics[width=\textwidth]{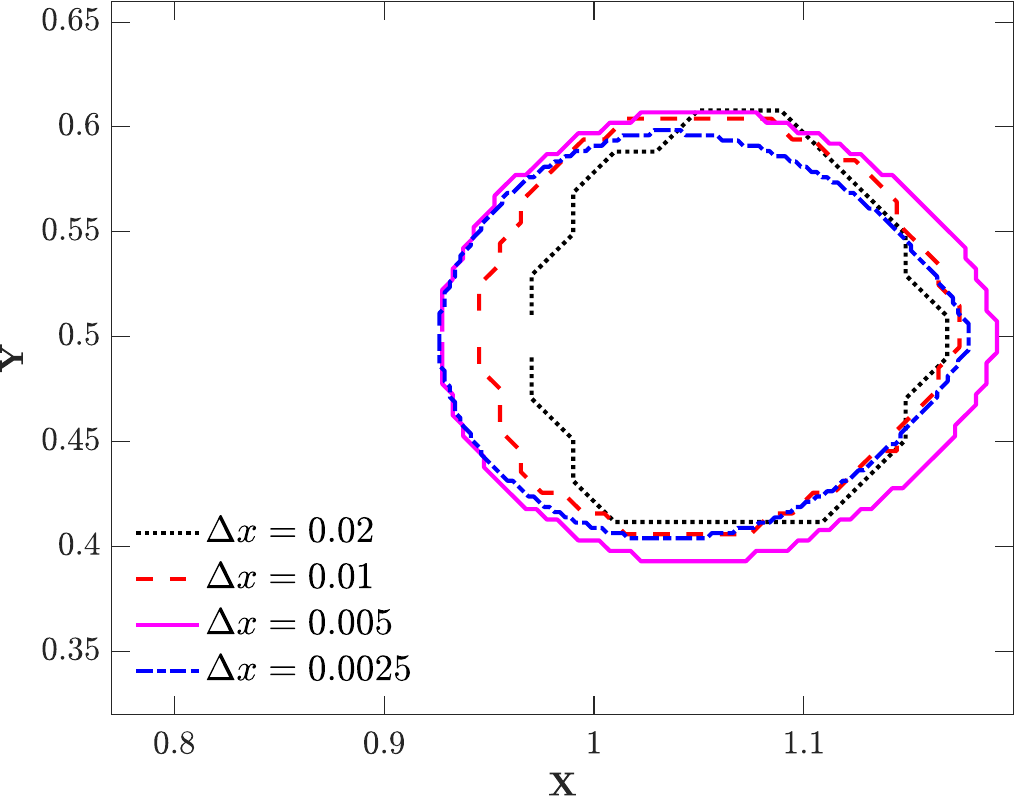}
      \end{center}
      \label{fig:err-case3}
     \end{subfigure}
     \hfill
     \begin{subfigure}[ht]{0.48\textwidth}
      \caption{}
      \begin{center}
        \includegraphics[width=\textwidth]{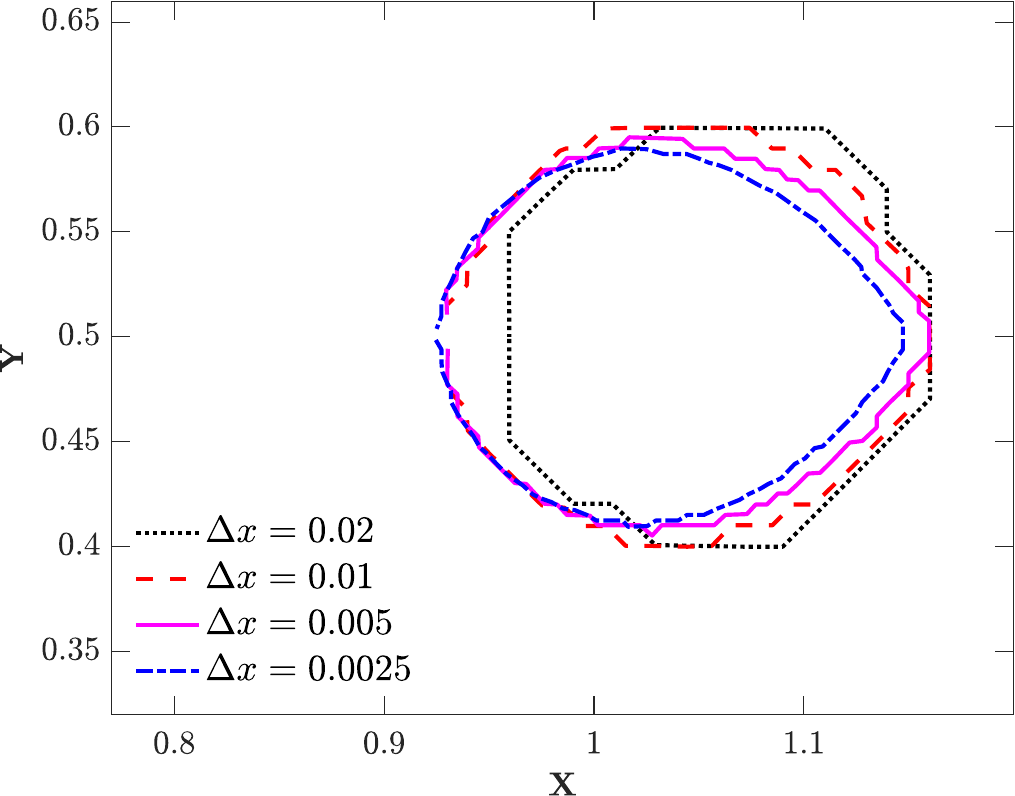}
      \end{center}
      \label{fig:err-case4}
     \end{subfigure}
     \caption{Grid shape convergence for different grid sizes at several transport regimes and different times. {(I) : $(Da, Pe, t, \Delta t)$ = (0.178, 6, 5,  $10^{-3}$), (II) : $(Da, Pe, t, \Delta t)$ = (178, 6, 0.75, $10^{-5}$), (III) : $(Da, Pe, t, \Delta t)$ = (178, 600, 5, $10^{-4}$), (IV) : $(Da, Pe, t, \Delta t)$ = (17800, 600, 6, $10^{-5}$) }}
     \label{fig:shape_convg}
\end{figure}

The grid size is chosen after a grid convergence study for the shape and volume of the grain. Figure \ref{fig:shape_convg} shows the grain shape convergence as the grid is refined. The required grid size for accurate simulation strongly depends on the Damk\"{o}hler and P\'{e}clet number. A more refined grid is needed for high Damk\"{o}hler and P\'{e}clet numbers. As apparent from  Figure \ref{fig:shape_convg}, the coarser grid size fails to even predict a physical shape particularly at high Pe and Da numbers. We chose $\Delta x = 0.005$ as the grid size for the simulations.

%

\begin{figure}[H]
     \centering
     \begin{subfigure}[ht]{0.48\textwidth}
     \caption{}
      \begin{center}
        \includegraphics[width=\textwidth]{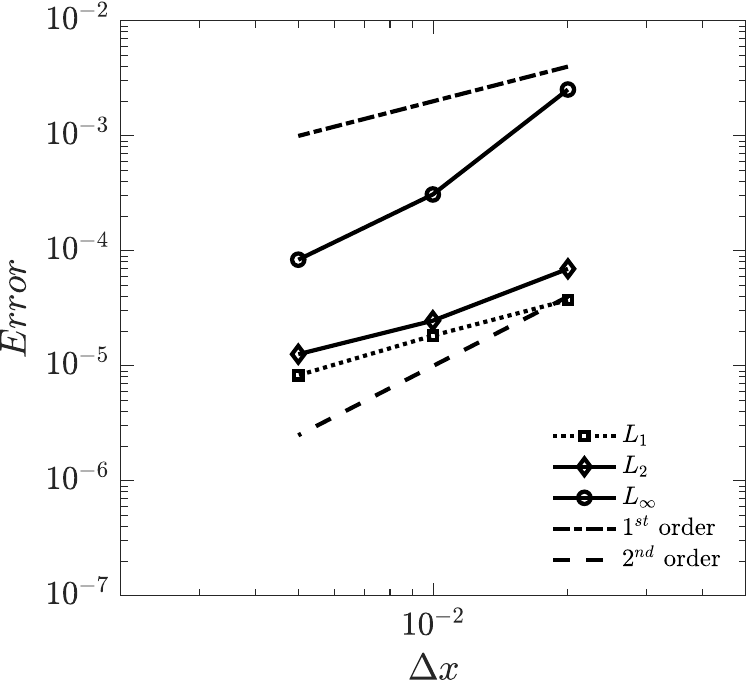}
      \end{center}
      \label{fig:grid-conv-err-case1}
     \end{subfigure}
     \hfill
     \begin{subfigure}[ht]{0.48\textwidth}
     \caption{}
      \begin{center}
        \includegraphics[width=\textwidth]{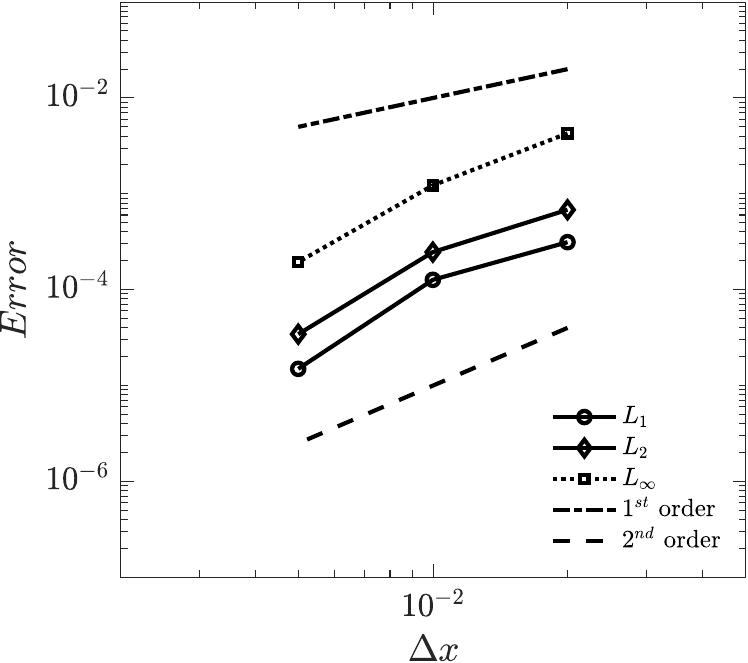}
      \end{center}
      \label{fig:grid-conv-err-case2}
     \end{subfigure}
     
     \bigskip
     \begin{subfigure}[ht]{0.48\textwidth}
      \caption{}
      \begin{center}
        \includegraphics[width=\textwidth]{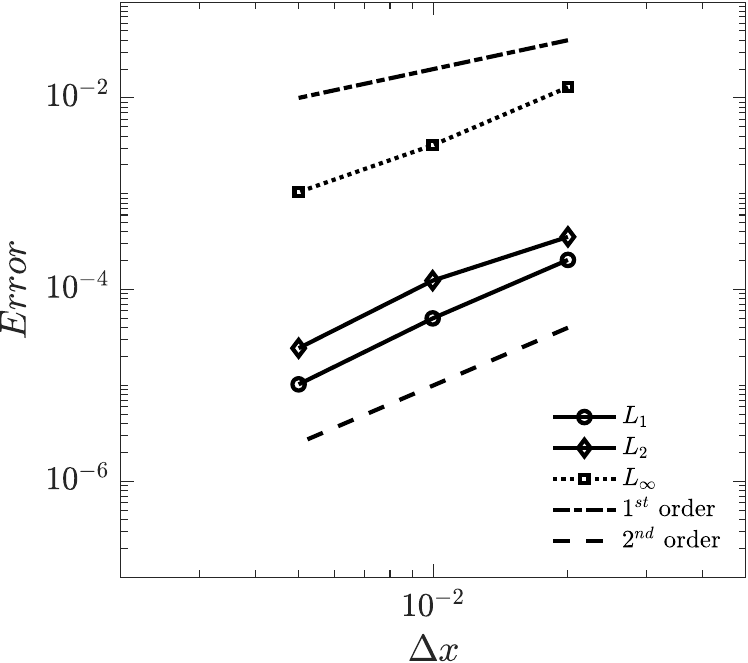}
      \end{center}
      \label{fig:grid-conv-err-case3}
     \end{subfigure}
     \hfill
     \begin{subfigure}[ht]{0.48\textwidth}
      \caption{}
      \begin{center}
        \includegraphics[width=\textwidth]{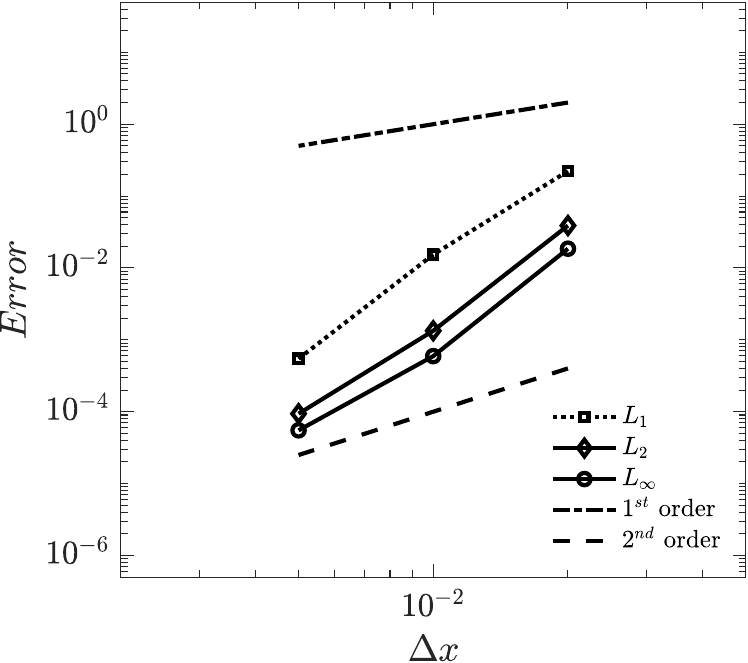}
      \end{center}
      \label{fig:grid-conv-err-case4}
     \end{subfigure}
     \caption{The grid convergence for {(I) : $(Da, Pe, \Delta t)$ = (0.178, 6, $10^{-3}$), (II) : $(Da, Pe, \Delta t)$ = (178, 6, $10^{-5}$), (III) : $(Da, Pe, \Delta t)$ = (178, 600, $10^{-4}$), (IV) : $(Da, Pe, \Delta t)$ = (17800, 600, $10^{-5}$) }}
     \label{fig:grid_conv}
\end{figure}

In Figures~\ref{fig:grid_conv}(\subref{fig:err-case1})~--~(\subref{fig:err-case4}), $L_1$, $L_2$ and $L_{\infty}$ relative errors of the grain volume (area) at various P\'{e}clet and Damk\"{o}hler numbers are plotted against the grid size. The errors are computed as the difference between the current and the successive refinement. {The convergence rate shows  second order of accuracy in space for all error norms, except for Case I which has first order of accuracy in space for $L_1$ and $L_2$ norms.} Figure 5 shows the symmetric difference error as a function of grid resolution $\Delta{x}$, following the method by~\cite{Ahn2007-ig}. The symmetric difference error $err_{sym}$ is computed as 
\begin{align}\label{eq:sym-diff-err}
err_{sym} = \left( V_{true} \cup V_{LS} \right) - \left( V_{true} \cap V_{LS}\right),
\end{align}
where $V_{true}$ and $V_{LS}$ refers to the true and level-set reconstructed areas or volumes. The convergence rate shows the order of accuracy is between first and second order.

\begin{figure}[H]
  \begin{center}
    \includegraphics[width=.5\textwidth]{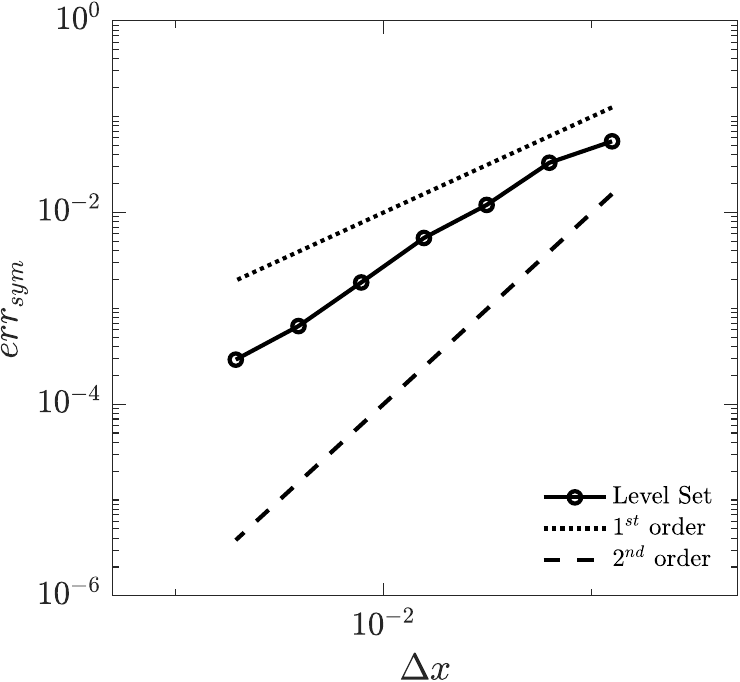}
  \end{center}
  \caption{The symmetric difference error for level-set {reconstruction} as a function of grid resolution {$\Delta{x}$ for $(Da, Pe, \Delta t)$ = (178, 600, $10^{-4}$)}.}
  \label{fig:sym-diff-error}
\end{figure}

\begin{figure}[H]
     \centering
     \begin{subfigure}[b]{0.5\textwidth}
     \caption{}
      \begin{center}
        \includegraphics[width=6cm, keepaspectratio]{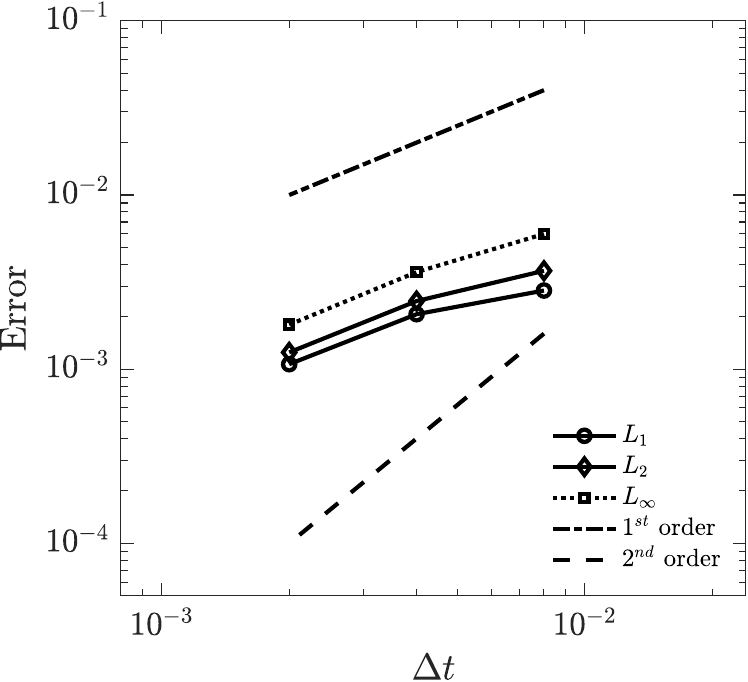}
      \end{center}
      \label{fig:err-time-step}
     \end{subfigure} 
      \\ 
      
     \begin{subfigure}[b]{0.5\textwidth}
     \caption{}
      \begin{center}
        \includegraphics[width=6cm, keepaspectratio]{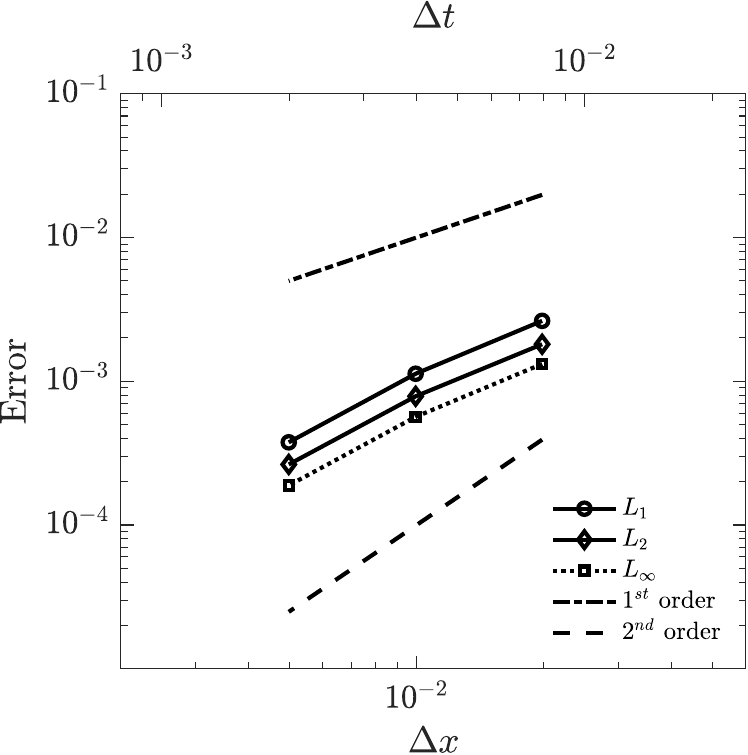}
      \end{center}
      \label{fig:err-time-step-grid}
     \end{subfigure}
  \caption{{(I) The time step convergence with $\Delta {x} = 0.005$ and (II) concurrent time step and grid size convergence} for $(Da, Pe)$ = (178, 600).}
  \label{fig:err-time-step-grid-all}
\end{figure}

In Figure~\ref{fig:err-time-step-grid-all}(\subref{fig:err-time-step}), we plot the volume convergence versus time step size for $Da=178$ and $Pe=600$. All error norms show first order of accuracy in time. This aligns with the first order time integration method that we have used. Although not implemented, the proposed GCIBM and LSIBM methods can be accompanied by higher order explicit temporal integration schemes to achieve higher temporal accuracy. {In Figure~\ref{fig:err-time-step-grid-all}(\subref{fig:err-time-step-grid}), we investigate the effects of volume convergence versus concurrent refinement of time step size and grid size. All error norms show first order of accuracy because the error is dictated by the first-order accurate time integration method.}


Next, the numerical stability of LSIBM is examined for flow  and reactive transport around  a dissolving circular grain. The test cases consist of a wide range of $Pe$ and $Da$ and are performed for multiple time steps ($\Delta t$).

The stability of LSIBM highly depends on the LS CFL condition mentioned in Table \ref{tab:timesteps}. In order to better test the stability, we set the initial concentration $C_{in}$ equal to 1 (i.e., $\hat C_{in} = C_c$). This ensures that the concentration at the interface is high, and the interface will move with a higher speed compared to cases with lower initial concentration.. The interface velocity is controlled by the  $Da/Pe$ ratio (as shown in  Table~\ref{tab:timesteps})) as well as the concentration near the interface. The $Da/Pe$ ratios for the test cases can be found in Table~\ref{tab:Da-Pe}. The cases with smaller $Da/Pe$ are more stable with respect to larger time steps (see, e.g., ($Da$, $Pe$) = (178,600) vs (178,0.6) in Table~\ref{tab:stab-cases-LSIBM}). For two cases with the same value of $Da/Pe$, the one with higher $Da$ is more stable at larger time steps (Table 4 ($Da$, $Pe$) = (17800,600) vs (178,6)). This can be explained as follows: for a fixed $Da/Pe$ ratio,  a low $Da$ corresponds to a negligible concentration gradient, i.e.  $C_{X'} \approx C_{X''}$  in Equation (22), and  $u_{\Gamma, Da\rightarrow0} \sim \frac{Da}{Pe} V_mC_c \left(3C_{X''}\right)$. For the same $Da/Pe$ ratio, but higher $Da$, a much larger concentration gradient is expected since the rate of consumption of solute due to reaction at the interface surpasses the supply of solute to the interface by diffusion and advection. This results in a steep concentration gradient between the solute concentration in proximity of the reacting surface and in the bulk, with a solute-depleted zone near the interface, i.e. $C_{X'} \approx 0$. Hence,  $u_{\Gamma, Da\rightarrow\infty} \sim \frac{Da}{Pe} V_mC_c C_{X''}$. This results in a lower  interface velocity for higher $Da$ numbers, when the $Da/Pe$ is fixed. The stability tests for multiple $Da$ and $Pe$ are listed in Table \ref{tab:stab-cases-LSIBM}. For each simulation, two time steps are reported, with the critical time step value for a stable solution in between such two values.

\begin{table}[H]
\begin{center}
\caption {Stability of LSIBM for different $Pe$, $Da$ and time steps ($\Delta x = 0.005$)} \label{tab:stab-cases-LSIBM} 
\begin{tabular}{l*{6}{c}r}
\hline
$\Delta t$ & $Da$ & $Pe$ & stable/unstable     \\
\hline
5E-4& $178$ & $6$ & stable  \\  
8E-4& $178$ & $6$ & unstable  \\
                            \hline
5E-5& $178$ & $0.6$ & stable \\
8E-5& $178$ & $0.6$ & unstable \\
\hline
1E-2& $178$ & $600$ & stable \\
4E-2& $178$ & $600$ & unstable \\
\hline
5E-3& $17800$ & $600$ & stable \\
3E-2& $17800$ & $600$ & unstable \\
\hline
1E-2& $1.78$ & $6$ & stable  \\
3E-2& $1.78$ & $6$ & unstable  \\
\hline
\end{tabular}
\end{center}
\end{table}

\subsubsection{Benchmark Validation and Dissolution Dynamics}

\begin{figure}[H]
  \begin{center}
    \includegraphics[clip, trim={0cm 1cm 1cm 1cm},width=0.9\textwidth]{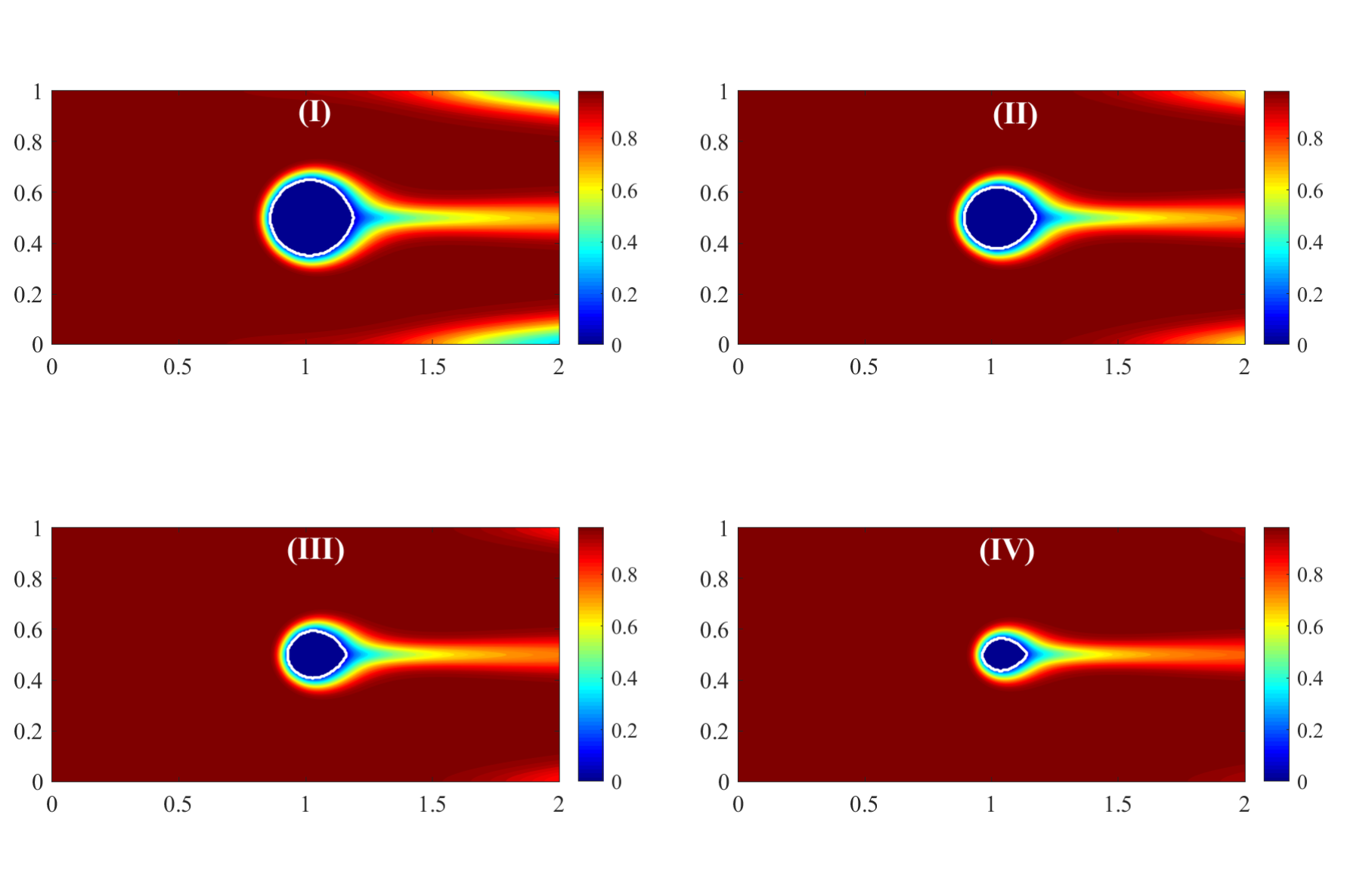}
  \end{center}
  \caption{The concentration for $(Da, Pe) = (178, 600)$ (Table~\ref{tab:Da-Pe} Case (V)) at times (I): $t = 3$, (II): $t = 4$, (III): $t = 5$, (IV): $t = 6$. The solid white line shows the grain interface. }
  \label{fig:concen}
\end{figure}

We now proceed with the validation of the LS-IBM algorithm against five solvers used
in the benchmark by \cite{molins2020simulation} (Chombo-Crunch~\cite{Molins2017-dx, Molins2014-pv, Molins2012-rk}, OpenFOAM-DBS~\cite{Soulaine2016-rd, soulaine2017mineral}, Lattice Boltzmann methods (LBM)~\cite{Yoon2012-oa}, vortex method~\cite{molins2020simulation, Gazzola2011-rx, Chatelin2016-pl, Chatelin2013-aj}, dissolFoam~\cite{Starchenko2016-eb}) . Figures \ref{fig:concen}(I)--(IV) show the evolution of the concentration as well as the grain interface  for $Da = 178$ and $Pe = 600$ (Table~\ref{tab:Da-Pe} Case (V)) at times, $t = 3, 4, 5, 6$, respectively. These overall dissolution patterns show a good agreement with the simulations in \cite{molins2020simulation} and \cite{soulaine2017mineral}.  

\begin{figure}
     \centering
     \begin{subfigure}[ht]{0.48\textwidth}
      \caption{}
      \begin{center}
        \includegraphics[width=\textwidth]{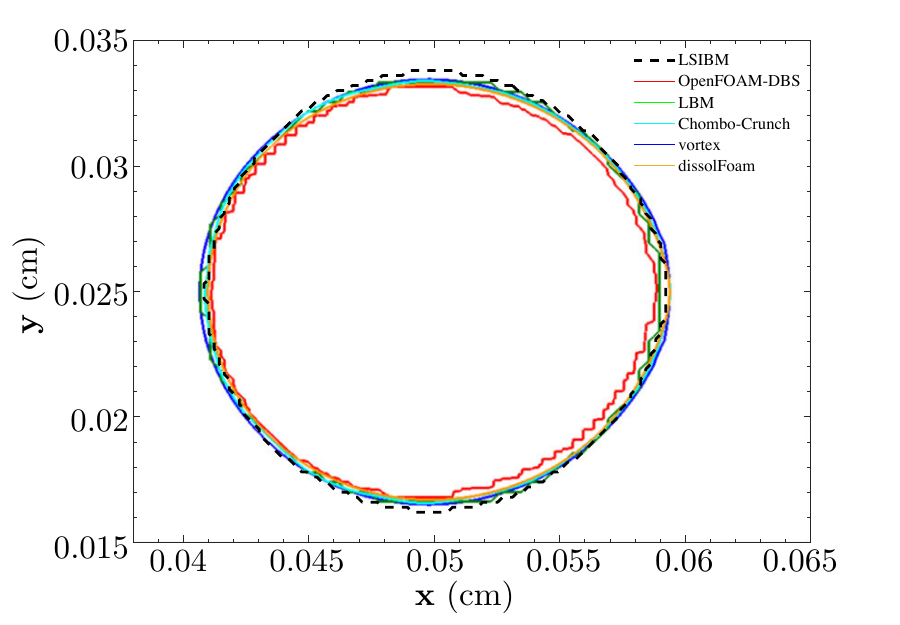}
      \end{center}
      \label{fig-t-15}
     \end{subfigure}
     \hfill
     \begin{subfigure}[ht]{0.48\textwidth}
      \caption{}
      \begin{center}
        \includegraphics[width=\textwidth]{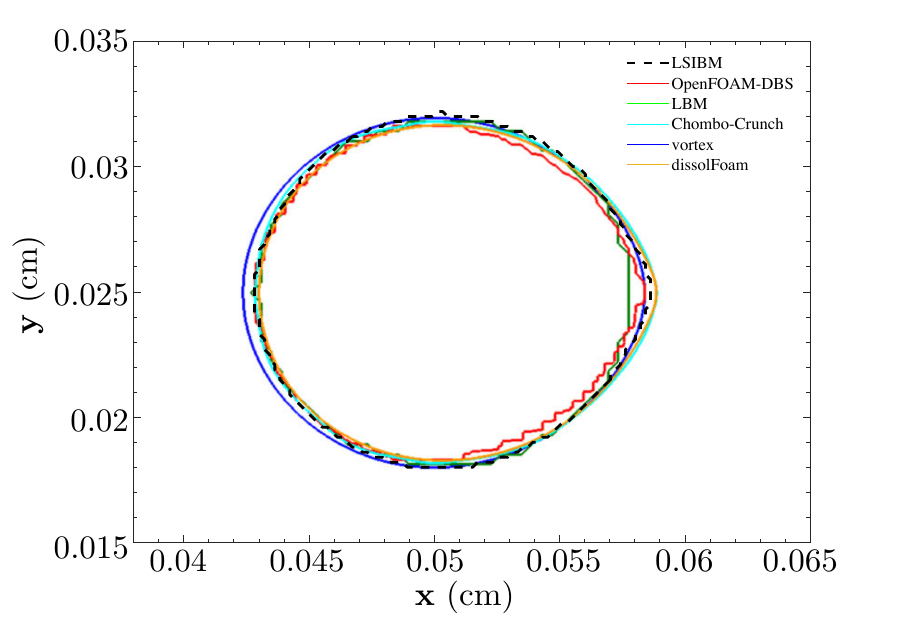}
      \end{center}
      \label{fig-t-30}
     \end{subfigure}
     
     \bigskip
     \begin{subfigure}[ht]{0.48\textwidth}
      \caption{}
      \begin{center}
        \includegraphics[width=\textwidth]{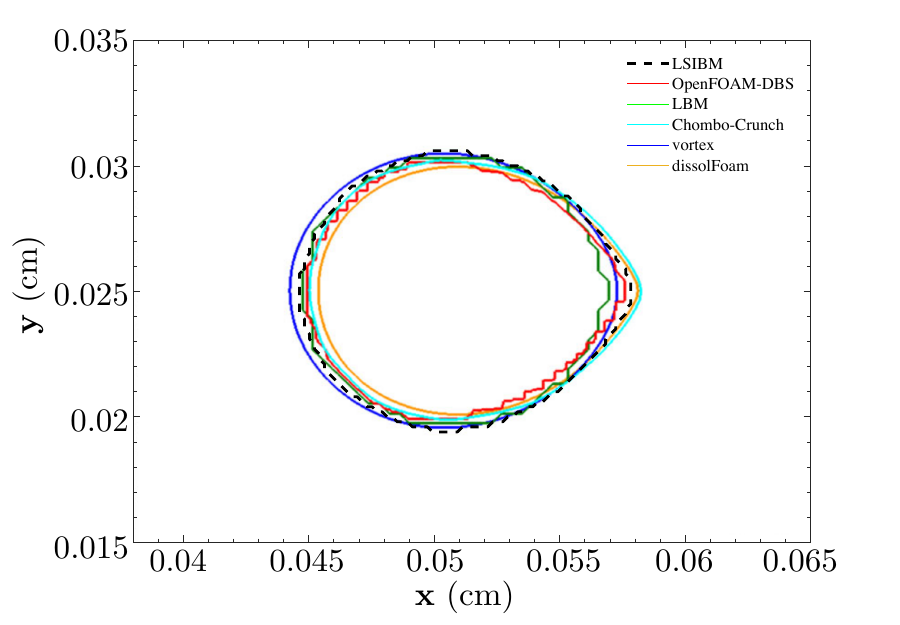}
      \end{center}
      \label{fig-t-45}
     \end{subfigure}
     \caption{The solid grain shape for $(Da, Pe) = (178, 600)$ at dimensional time $t = $ (I) $15$, (II) $30$, and (III) $45$ minutes from different methods.}
     \label{fig:da_pe_178_600_time}
\end{figure}

In Figures~\ref{fig:da_pe_178_600_time}(\subref{fig-t-15})--(\subref{fig-t-45}), the grain shape for $Da = 178$ and $Pe = 600$ (Table~\ref{tab:Da-Pe} Case (V)) is plotted along with the results in \cite{molins2020simulation} for a more quantitative comparison.  Due to the non-symmetric concentration field along the $x$-axis, the grain surface evolves in a non-uniform manner. The diamond (petal) shape of the solid grain is a result of higher dissolution rate at the  upstream side (far left)  of the grain, while the downstream (far right) side of the grain is exposed to lower acid concentration, dissolves with a lower rate and its boundary moves much less. The grain dissolution pattern is close and in good agreements with the benchmark results. Our results show a closer match to the results from Chombo-Crunch. This is due to the fact that both methods are based on an Eulerian finite-volume framework.

\begin{figure}
     \centering
     \begin{subfigure}[b]{0.48\textwidth}
      \caption{}
      \begin{center}
        \includegraphics[width=\textwidth]{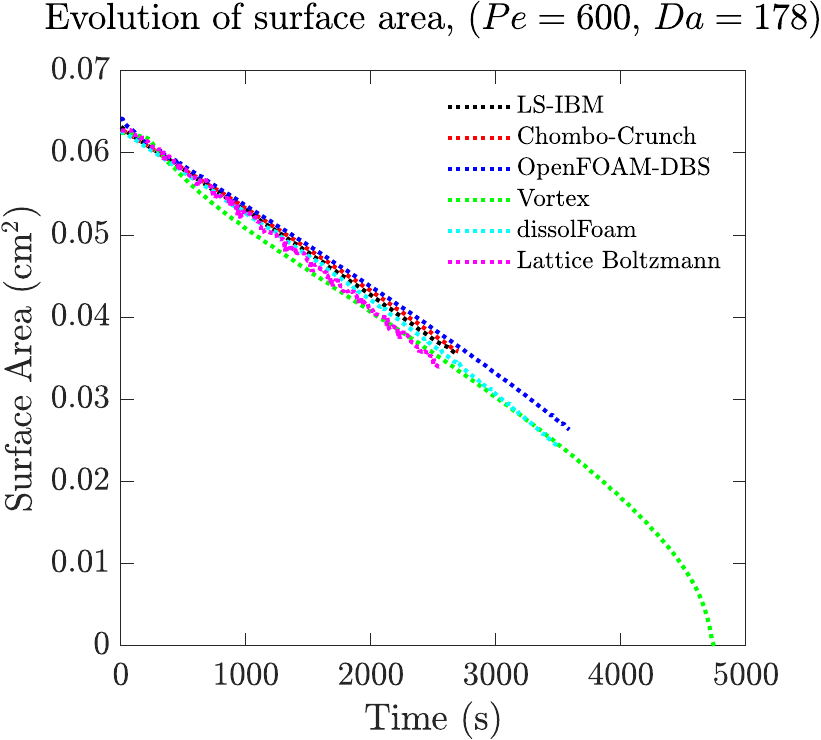}
      \end{center}
      \label{fig:surfaceareabase}
     \end{subfigure}
     \hfill
     \begin{subfigure}[b]{0.48\textwidth}
      \caption{}
      \begin{center}
        \includegraphics[width=\textwidth]{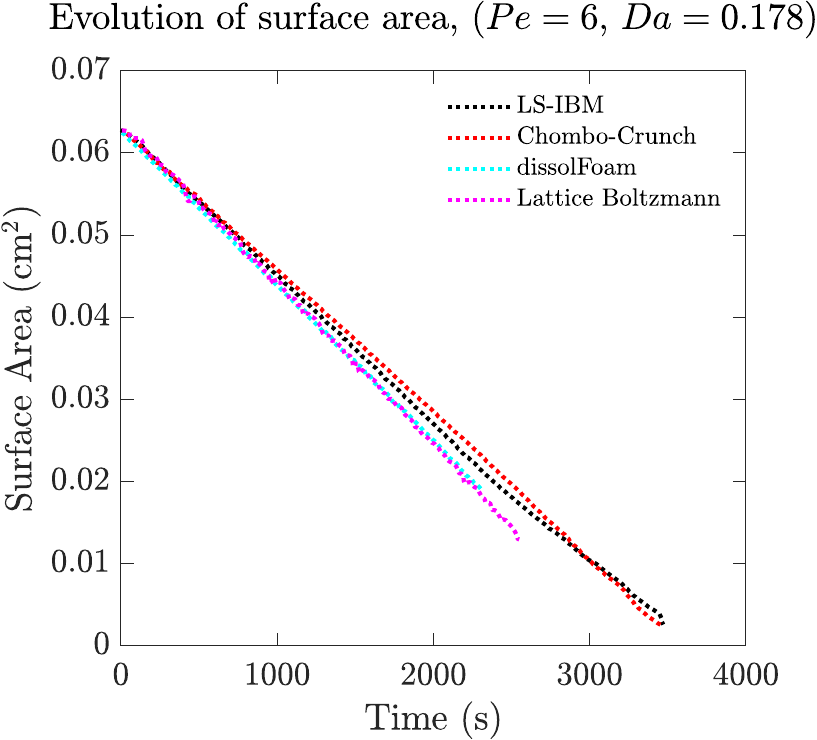}
      \end{center}
      \label{fig:surfaceareauniform}
     \end{subfigure}
     \caption{Grain surface area temporal evolution for (I) $Pe=600$ and $Da=178$ and (II) $Pe=6$ and $Da=0.178$: comparison between LS-IBM and the methods in
     \cite{molins2020simulation}.}
     \label{fig:surfacearea_all}
\end{figure}

Additionally, we have calculated the evolution of the total surface area (perimeter in 2D) of the calcite grain with time.
In Figures~\ref{fig:surfacearea_all}(\subref{fig:surfaceareabase}) and~(\subref{fig:surfaceareauniform}),   we show simulation results corresponding to two scenarios with $Da = 178$, $Pe = 600$ (Table~\ref{tab:Da-Pe} Case (V)) and $Da = 0.178$, $Pe = 6$ (Table~\ref{tab:Da-Pe} Case (I)). The surface evolution for both cases is in agreement  with the results presented in  \cite{molins2020simulation} where five different numerical algorithms were compared. Again, our results and those reported by  Combo-Crunch are closer to each other than other methods. This is expected because   both methods use a finite volume scheme along with level-set to update the geometry. 

\begin{figure}[H]
  \begin{center}
    \includegraphics [clip, trim={0cm 0cm 0cm 0cm},width=1\textwidth]{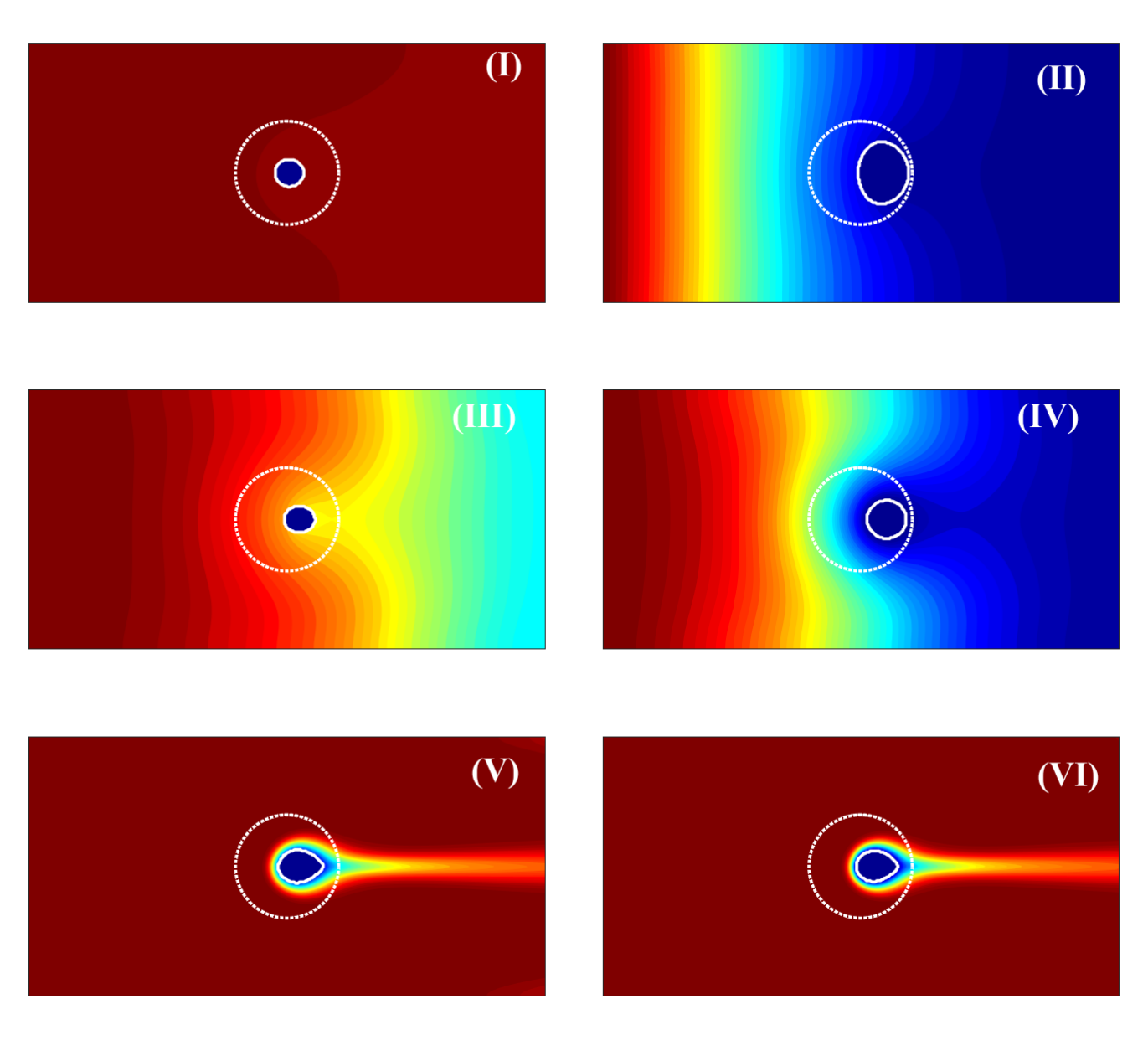}
      \caption{The concentration contour for the 6 simulation cases (I, II, III, IV, V, and VI) of   Table \ref{tab:Da-Pe} at different time snapshots ($t_I = 6$, $t_{II} = 0.2$, $t_{III} = 1.5$, $t_{IV} = 0.9$, $t_{V} = 6$ and $t_{VI} = 8$). The solid and dotted white lines show the evolved and initial grain surface, respectively. }
        \label{fig:diff-da-pe}
  \end{center}
\end{figure}

 When $Da = 178$ and $Pe =600$ (Table~\ref{tab:Da-Pe} Case (V)) , the grain starts changing to a petal shape from its initial circular shape during  dissolution. This is due to the lower acid concentration at the stagnation point on the right side of the cylinder and thus, lower local dissolution rate at that location (see Figure \ref{fig:diff-da-pe}(V)). For  $Da = 0.178$ and $Pe =6$ (Table~\ref{tab:Da-Pe} Case (I)),  the grain shape remains circular throughout the dissolution process.

The simulation results and the grain shape patterns agree with the $Pe/Da$ diagram for dissolution regimes reported in \cite{soulaine2017mineral}. Similar to \cite{soulaine2017mineral}, three distinct grain shape patterns are observed. When the reaction rate is small (i.e. $Da \le 1$, Table~\ref{tab:Da-Pe} Case (I)) the grain dissolves uniformly, the grain center almost stays unchanged and the initial cylindrical shape is preserved (Fig.~\ref{fig:diff-da-pe} I). However for larger $Da$ (Table~\ref{tab:Da-Pe} Cases (II)-(VI)), the grain shape starts deviating from its initial cylindrical shape and the grain center of mass moves due to non-uniform dissolution. For moderate to low P\'{e}clet number ($Pe \le 10$, Table~\ref{tab:Da-Pe} Cases (II)-(IV)) the grain surface stays smooth (Fig.~\ref{fig:diff-da-pe} II -IV). Differently from~\cite{soulaine2017mineral}, we observe three distinct sub-regimes. For very small P\'{e}clet numbers ($Pe \le 1$), the grain elongation is perpendicular to the flow direction (Fig.~\ref{fig:diff-da-pe} II, Table~\ref{tab:Da-Pe} Case (II)): because of the low P\'{e}clet number, the right side of the grain is exposed to very small amount of acid ions, and since the $Da$ is not very large the reaction rate near the top and bottom of the grain is smaller when compared to Case (IV) (Table~\ref{tab:Da-Pe}).Therefore, low dissolution rates at the top and bottom, and high rate on the left side of the grain, cause a horizontally skewed shape. As the Pe increases, the right side of the grain is exposed to more reactive acid ions and an elongation in the flow direction is observed (Table~\ref{tab:Da-Pe} Case (III)). However, this elongation becomes less apparent when the Damk\"{o}hler is very large ($Da \ge 10$, Table~\ref{tab:Da-Pe} Case (IV) ). Finally for large P\'{e}clet numbers, the diamond (petal) pattern of the grain (Fig.~\ref{fig:diff-da-pe} V and VI, Table~\ref{tab:Da-Pe} Case (V) and (VI)) corresponding to the third regime in~\cite{soulaine2017mineral}, is observed.

\begin{figure}[H]
  \begin{center}
    \includegraphics [clip, trim={2cm 0cm 0cm 0cm},width=1.1\textwidth]{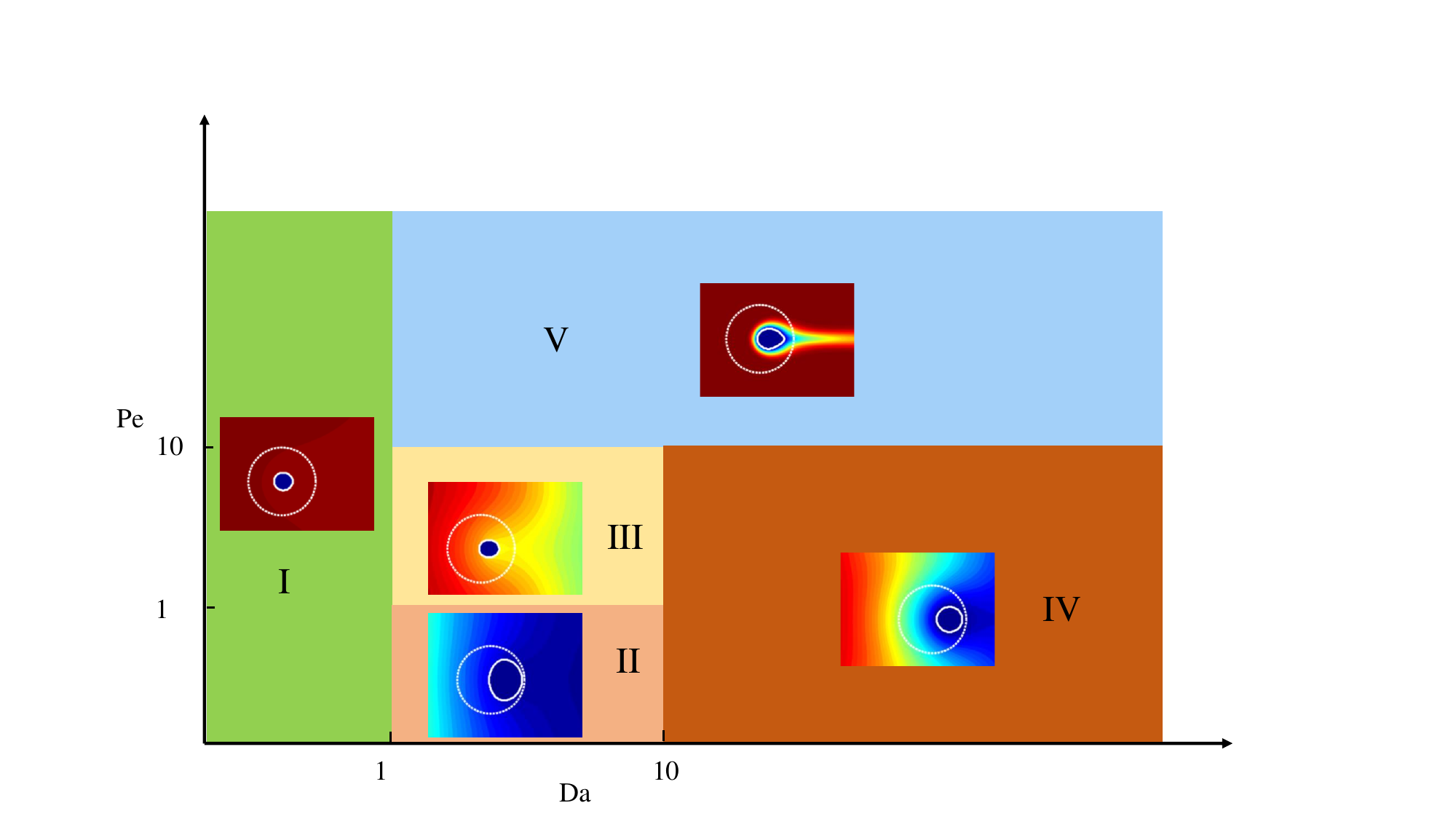}
      \caption{The concentration contour and grain shape for different values of $Da$ and $Pe$ as listed in  Table \ref{tab:Da-Pe} at different time snapshots ($t_I = 6$, $t_{II} = 0.2$, $t_{III} = 1.5$, $t_{IV} = 0.9$, $t_{V} = 6$ and $t_{VI} = 8$). The solid and dotted white lines show the evolved and initial grain surface respectively. }
        \label{fig:phase-da-pe}
  \end{center}
\end{figure}

In Figure \ref{fig:phase-da-pe}, we show the $Da$--$Pe$ regime diagram for the grain shape, which includes the additional regimes discussed above.  We have summarized the characteristics of each regime and sub-regime in Table \ref{tab:Da-Pe-phase}.

\begin{table}[H]
\begin{center}
\caption {Dissolution regime characteristics} \label{tab:Da-Pe-phase} 
\begin{tabular}{l*{6}{c}r}
\hline
Regime & grain center & grain shape   \\
\hline
(I) & unchanged & cylindrical \\
(II) & moves to the left & horizontally skewed \\
(III) & moves to the left & vertically skewed \\
(IV) & moves to the left & cylindrical \\
(V) & moves to the left & diamond (petal) \\
\hline
\end{tabular}
\end{center}
\end{table}

\subsection{Diffusion limited precipitation and unstable dendritic growth}\label{sec:results3}
\begin{figure}[H]
  \begin{center}
    \includegraphics[width=0.5\textwidth]{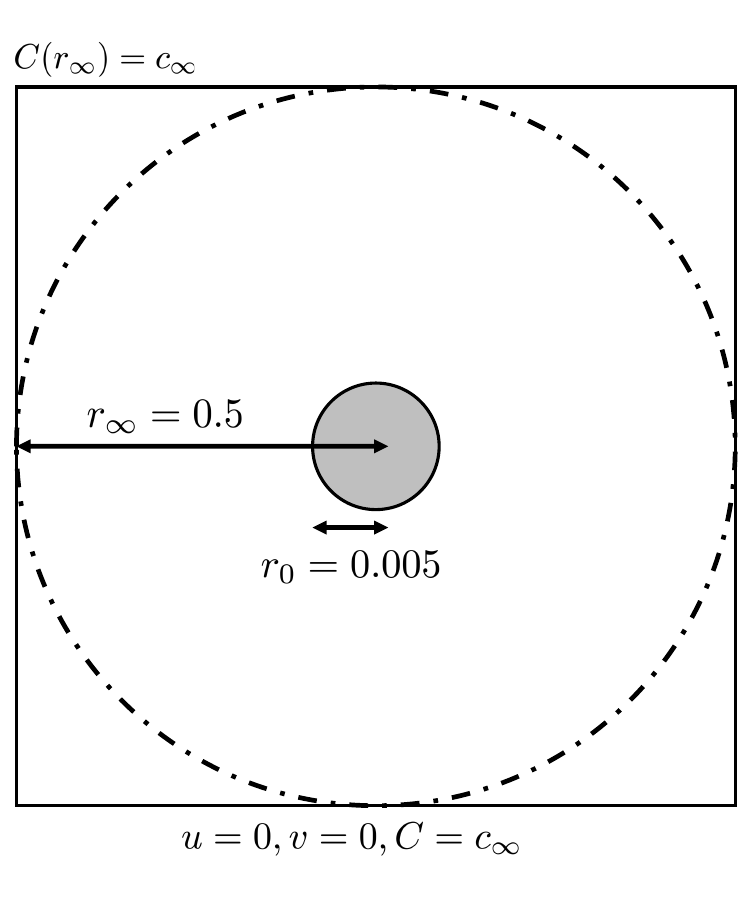}
  \end{center}
  \caption{Computational domain and BCs for unstable dendritic growth.}
  \label{fig:testdomain2}
\end{figure}

In order to further investigate the capability of the proposed LS-IBM to handle complex interface dynamics, we compute both the symmetric and unstable dendritic growth of a small circular nucleus with initial radius  $r_0$ (Figure~\ref{fig:testdomain2}). The growth occurs due to the surface precipitation of a solute with initial value of $c_\infty$. The dynamic of the solid-fluid interface and resulting growth pattern can be anticipated as a moving boundary (Stefan problem) \cite{meakin1998fractals, meakin1984monte, xu2011phase}. The Damk\"{o}hler number of the problem is high and therefore the transport is under diffusion limited condition.

The crystal growth problem due to diffusion limited precipitation is extensively studied in the reactive transport research community. Lattice Boltzmann method \cite{kang2004lattice}, smoothed particle hydrodynamics (SPH) \cite{tartakovsky2007simulations}, level set method \cite{li2008level} and phase field method \cite{xu2011phase} have all been used, among others, to model  stable and unstable crystal growth. 

\begin{figure}[H]
\vspace{-2cm}
  \begin{center}
    \includegraphics [clip, trim={3cm 0cm 6.6cm 0cm},width=1\textwidth]{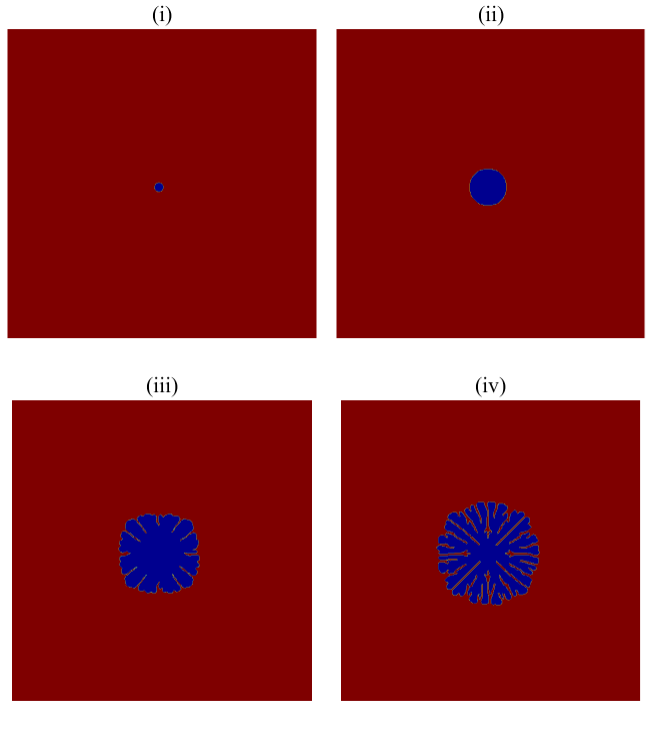}
      \caption{The growth pattern for $c_\infty = 1$ at time $t = 0.001$ at different Damk\"{o}hler numbers: 
      (i) $Da = 10$, (ii) $Da = 10^2$, (iii) $Da = 10^3$, (iv) $Da = 10^4$. }
      \label{fig:dengrowth-cinf1-t1}
  \end{center}
  
\end{figure}

\begin{figure}[H]
\vspace{-1cm}
  \begin{center}
    \includegraphics [clip, trim={1cm 0cm 3cm 0cm},width=1\textwidth]{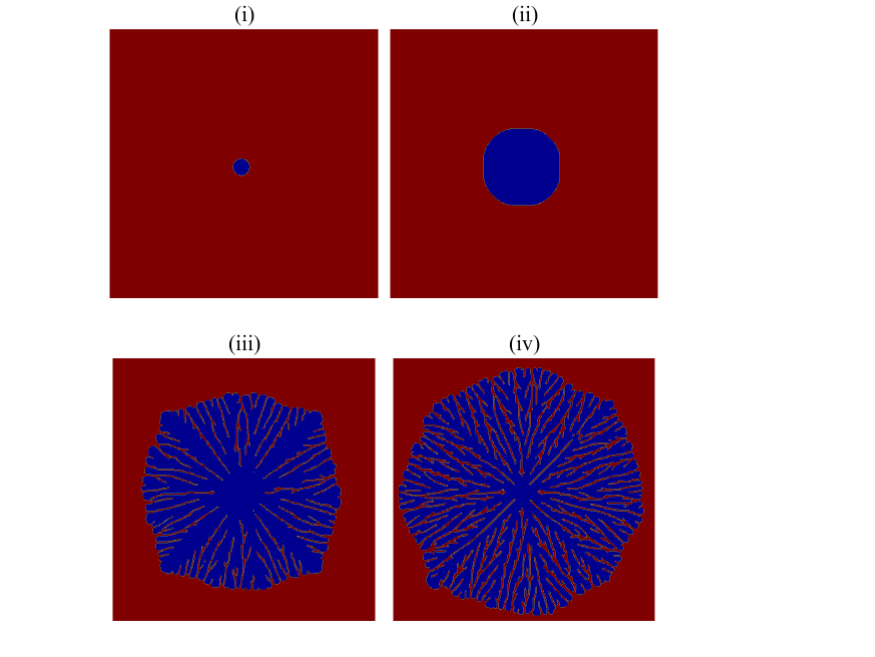}
      \caption{The growth pattern for $c_\infty = 1$ at time $t = 0.003$ at different Damk\"{o}hler numbers: 
      (i) $Da = 10$, (ii) $Da = 10^2$, (iii) $Da = 10^3$, (iv) $Da = 10^4$.  }
      \label{fig:dengrowth-cinf1-t3}

  \end{center}
\end{figure}

We consider the initial radius of a solid nucleus to be $r_0 = 0.005$, placed in a unit square domain. The concentration at radius $r_\infty = 0.5$ is held constant and equal to the initial concentration $c(r_\infty) = c_\infty$. Two values for the initial concentrations $c_\infty = \{0.1, 1\}$ and four values for Damk\"{o}hler number $Da = \{10, 100, 1000, 10000\}$ are considered. The domain is discretized in a 600 $\times$ 600 uniform grid. The growth patterns for $c_\infty = 1$ at two different instances in time  are shown in Figures \ref{fig:dengrowth-cinf1-t1}  and \ref{fig:dengrowth-cinf1-t3}. The effect of Damk\"{o}hler number is depicted in each Figure.  The process is more reaction-limited at smaller Damk\"{o}hler numbers ($Da \approx O(1)$) and results in a stable uniform growth with a smooth interface, see Figures \ref{fig:dengrowth-cinf1-t1}(i) and \ref{fig:dengrowth-cinf1-t3}(i). As the Damk\"{o}hler number increases, the diffusion rate becomes less and less adequate to sustain a uniform concentration around the boundary and the interface becomes increasingly less smooth, see Figures \ref{fig:dengrowth-cinf1-t1}(ii) and \ref{fig:dengrowth-cinf1-t3}(ii). Ultimately, at very high Damk\"{o}hler numbers ($Da = 10^3, 10^4)$, the reaction rate is much faster than the diffusion rate  and the process is  diffusion limited: as a result, any small perturbation will lead to a dendritic growth pattern, see Figures \ref{fig:dengrowth-cinf1-t1}(iii, iv) and \ref{fig:dengrowth-cinf1-t3}(iii, iv). The growth of a solid crystal by super saturated solute precipitation for high reaction rates is governed by the Mullins-Sekerka instability~ \cite{mullins1964stability}.

\begin{figure}[H]
  \begin{center}
    \includegraphics [clip, trim={1cm 0cm 3cm 0cm},width=1\textwidth]{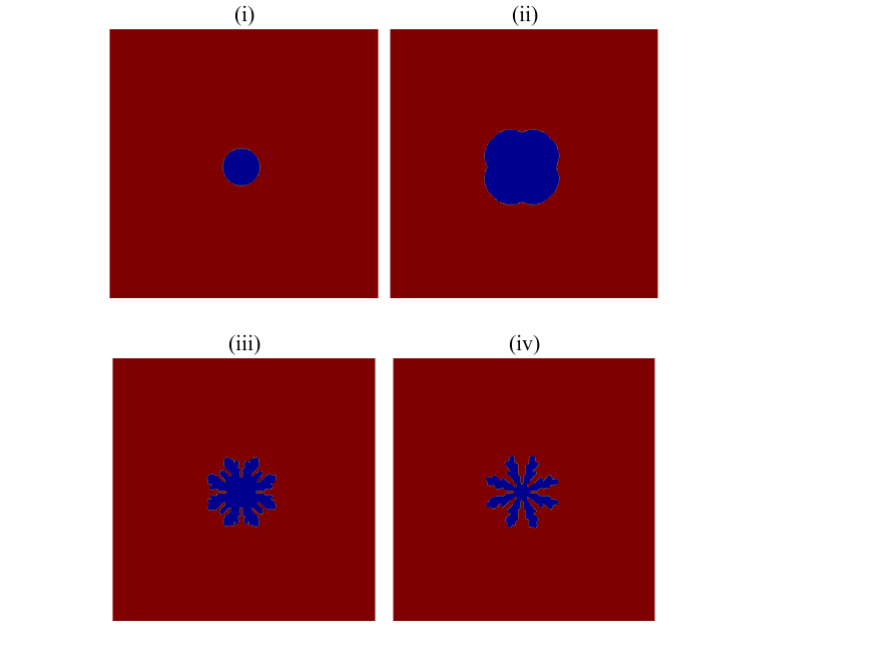}
      \caption{The growth pattern for $c_\infty = 0.1$ at time $t = 0.1$ at different Damk\"{o}hler numbers: 
      (i) $Da = 10$, (ii) $Da = 10^2$, (iii) $Da = 10^3$, (iv) $Da = 10^4$. }
        \label{fig:dengrowth-cinf0-1-t1}
  \end{center}
\end{figure}

\begin{figure}[H]
  \begin{center}
    \includegraphics [clip, trim={1cm 0cm 3cm 0cm},width=1\textwidth]{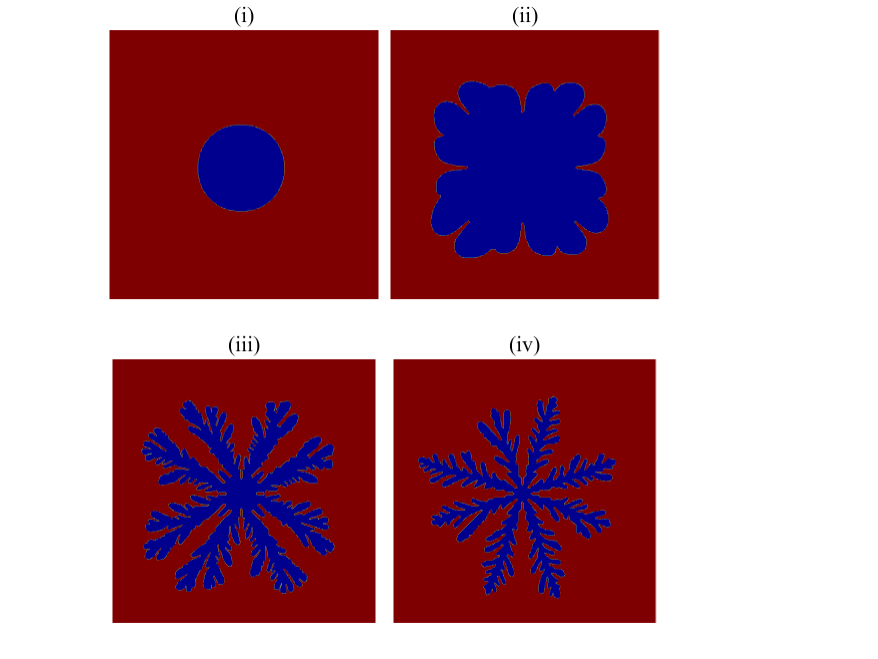}
      \caption{The growth pattern for $c_\infty = 0.1$ at time $t = 0.3$ at different Damk\"{o}hler numbers: 
      (i) $Da = 10$, (ii) $Da = 10^2$, (iii) $Da = 10^3$, (iv) $Da = 10^4$. }
        \label{fig:dengrowth-cinf0-1-t3}
  \end{center}
\end{figure}

In order to investigate the effects of $c_\infty$ on the growth pattern, we have simulated the problem at $c_\infty = 0.1$. The results for various Damk\"{o}hler numbers at two time snapshots are presented in Figures \ref{fig:dengrowth-cinf0-1-t1} and \ref{fig:dengrowth-cinf0-1-t3}. A more branched growth compared to the case with $c_\infty$ is evident. The branches perish as they advance to the far field due to the competition with their adjacent branches to absorb mass. The lack of initial mass stops the growth of branches and unlike the previous case, the pattern will have thinner branches.

\begin{figure}[H]
  \begin{center}
    \includegraphics [width=0.5\textwidth]{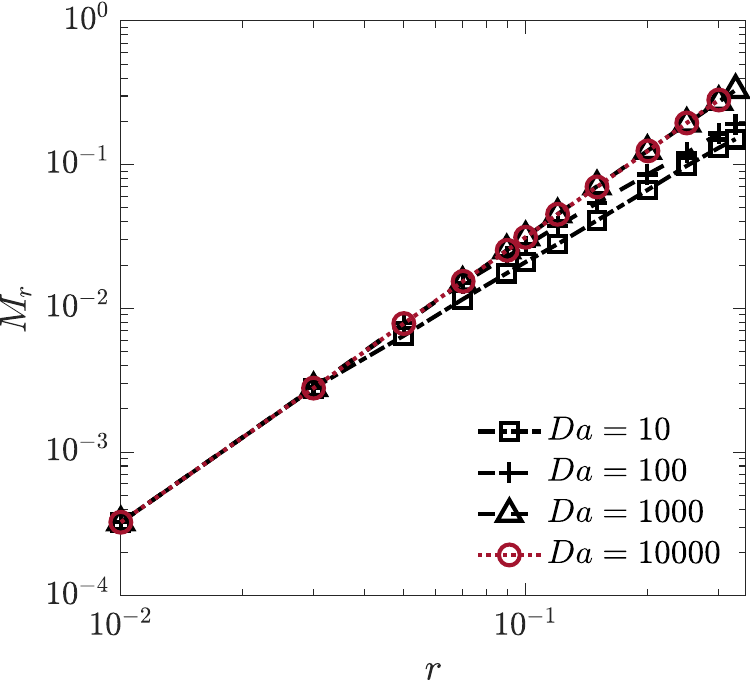}
      \caption{$M_r$ vs. $r$ for dendritic growth for $c_\infty = 0.1$}
        \label{fig:Mr-r}
  \end{center}
\end{figure}

Further, we compare the  effective fractal dimension obtained in the previous simulations with those in the literature. The effective fractal dimension, $D_f$, characterizes the volume (or area in 2D) of the crystal within a radius $r$ and can be determined through the power law relationship \eqref{eq:LSfractal}
\begin{align}\label{eq:LSfractal}
    M_r \sim r^{D_f},
\end{align}
where $M_r$ is the volume (area in 2D) of the solid. We use the box-counting method to measure the fractal dimension for different Damk\"{o}hler numbers.  In Figure~\ref{fig:Mr-r} we plot the area of the solid $M_r$ versus the radius $r$ in a logarithmic scale for different Damk\"{o}hler numbers. As expected from Equation~\eqref{eq:LSfractal}, $\log(M_r)$ increases linearly with $\log(r)$. 
The effective fractal dimensions are reported in Table \ref{tab:DfComp}. The fractal dimension measured by SPH method \cite{tartakovsky2007simulations}, LBM \cite{kang2004lattice}, large scale off-lattice diffusion limited aggregation (DLA) model \cite{meakin1998fractals} and level set method of \emph{Li et  al.} \cite{li2008level} at extremely high reaction rate ($Da = O(10^3 \sim 10^4)$) are $D_f = 1.66$, $1.715$ and $1.75$ are $1.82$, respectively.   Our results, summarized in Table \ref{tab:DfComp},  are in close agreement with the previously cited values. In most cases, we want to trigger these instabilities anyway, so perturbations induced by the grid are not necessarily undesirable.
\begin{table}[H]
\begin{center}
\caption {Comparison of fractal dimension $D_f$ for dendritic growth with different Damk\"{o}hler numbers} \label{tab:DfComp} 
\begin{tabular}{l*{5}{c}r}
\hline
$Da$ & 10 &100 &1000 &10000 \\
\hline
Kang et. al (2004) \cite{kang2004lattice} & - & - &$1.715$ & -\\
Kang et. al (2004) \cite{kang2004lattice} & $2.0$ & $1.88$ &$1.75$ & -\\
Tartakovsky et. al (2007) \cite{tartakovsky2007simulations} & $2.0$ & $1.74$ &$1.66$ & -\\
Li et. al (2008) \cite{li2008level} & $2.0$ & $2.0$ &$1.85$ & $1.82$\\
Current & $2$ & $1.99$ & $1.81$& $1.74$\\
\hline
\end{tabular}
\end{center}
\end{table}

\begin{figure}[H]
  \begin{center}
    \includegraphics [clip, trim={5cm 6cm 5cm 4cm},width=1\textwidth]{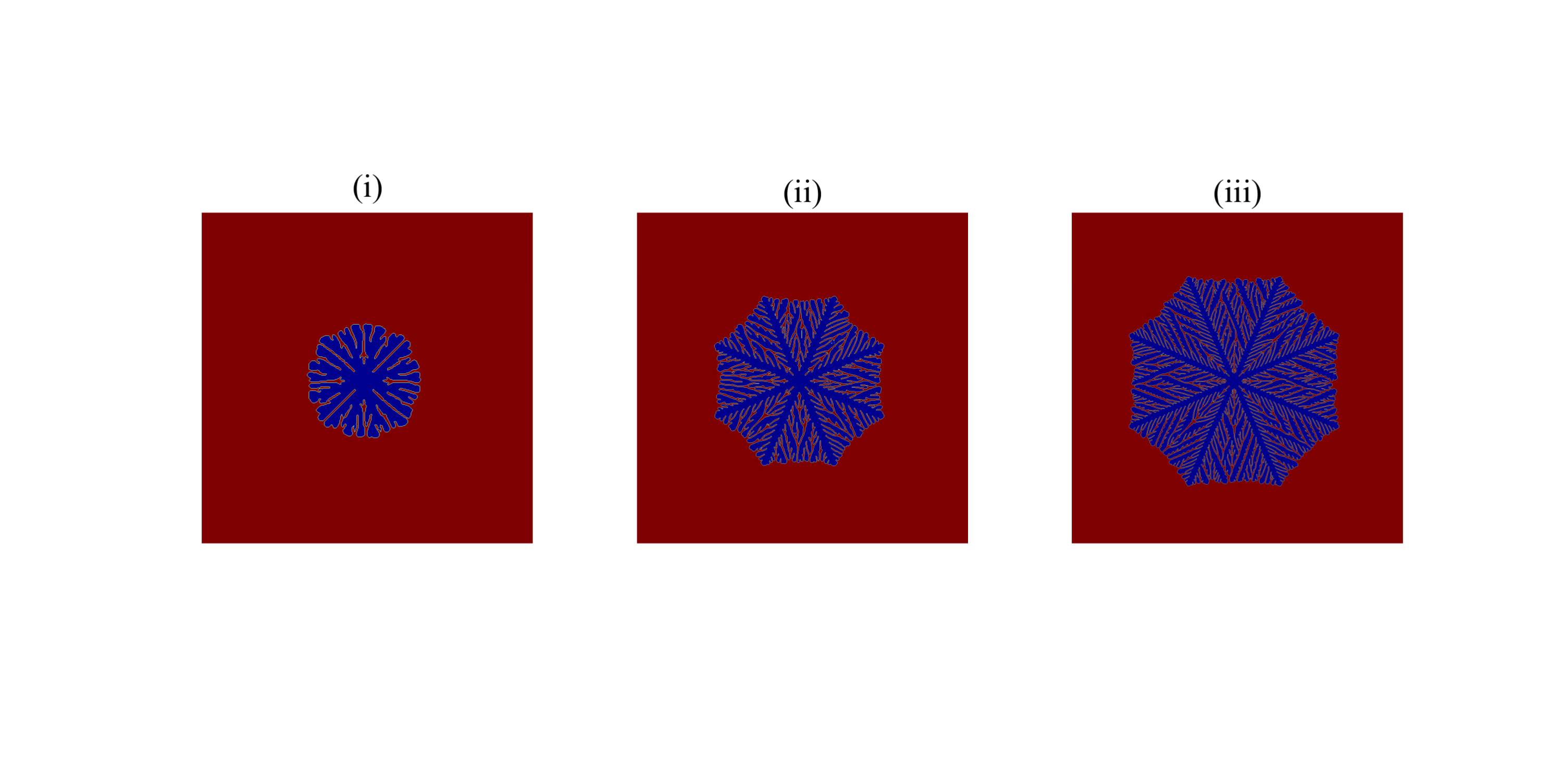}
      \caption{The growth pattern for $c_\infty = 1$ and $Da = 10^4$ {with $\Delta{t} = 10^{-7}$} at time $t = 0.001$ at different grid resolutions: 
      (I) $\Delta x = \dfrac{1}{400}$, (II) $\Delta x = \dfrac{1}{600}$ and (III) $\Delta x = \dfrac{1}{800}$. }
        \label{fig:dengrowth-comp}
  \end{center}
\end{figure}

\begin{figure}[H]
  \begin{center}
    \includegraphics [width=0.5\textwidth]{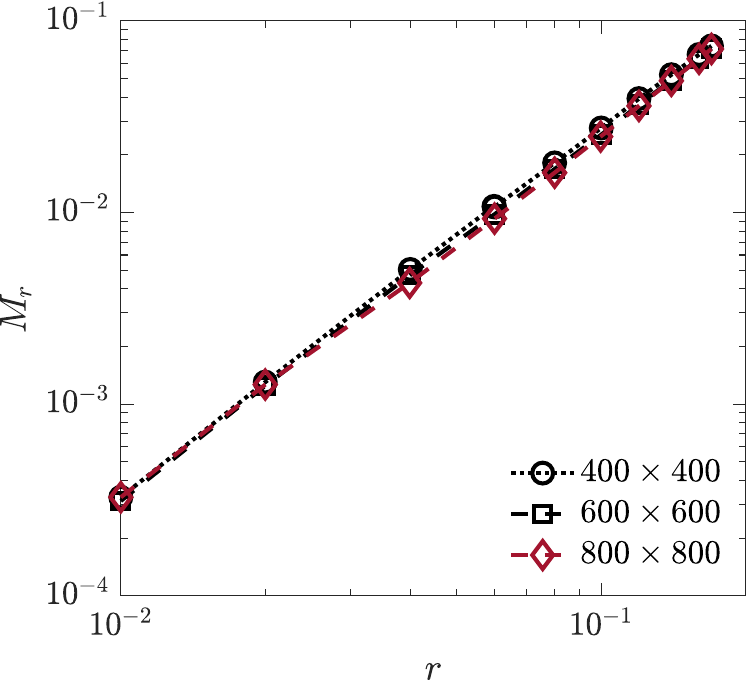}
      \caption{$M_r$ vs. $r$ for dendritic growth for $c_\infty = 1$ and $Da = 10^4$ at three different grid resolutions {with $\Delta{t} = 10^{-7}$}.}
        \label{fig:Mr-r-comp}
  \end{center}
\end{figure}

In the previous simulations, we use a $600\times 600$ grids. As observed by  Juric and Tryggvason \cite{juric1996front}, the grid size has a noticeable effect on the solution. To study the impact of the grid size on the solution and fractal dimension, we conduct simulations for $Da=10000$ and three different grid resolutions: $\Delta x=1/400$, $\Delta x=1/600$ and $\Delta x=1/800$. For the largest grid size the interface main branches are greater in numbers (16) compared to higher resolution solutions (8 main branches). On the other hand the two higher resolution solutions are more similar and each have 8 large branches. While the solutions in  Figures \ref{fig:dengrowth-comp} (ii, iii) are not fully converged in the sense that the two solutions are completely identical,  the physical phenomena are fully resolved for both grids since  the fractal dimension and the overall trend remains unchanged as shown in Figure \ref{fig:Mr-r-comp}. Furthermore, starting from an initially perturbed (non circular) interface will reduce such differences resulting from the grid size. These observations are consistent with those by   \cite{juric1996front}.

\begin{figure}[H]
     \centering
     \begin{subfigure}[ht]{0.48\textwidth}
     \caption{}
      \begin{center}
        \includegraphics[clip, trim={3cm 1cm 3cm 1.0cm},width=\textwidth]{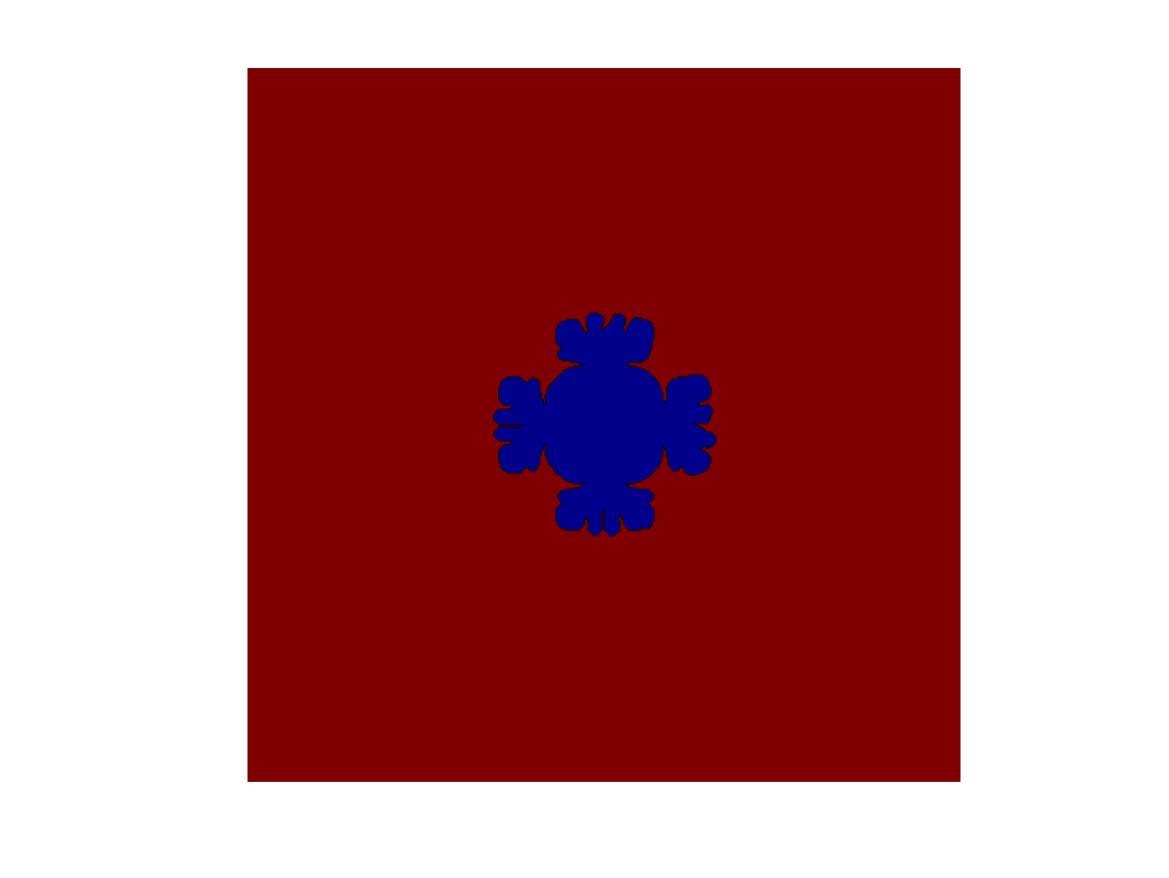}
      \end{center}
      \label{fig:s_1000}
     \end{subfigure}
     \hfill
     \begin{subfigure}[ht]{0.48\textwidth}
      \caption{}
      \begin{center}
        \includegraphics[clip, trim={3cm 1cm 3cm 1.0cm},width=\textwidth]{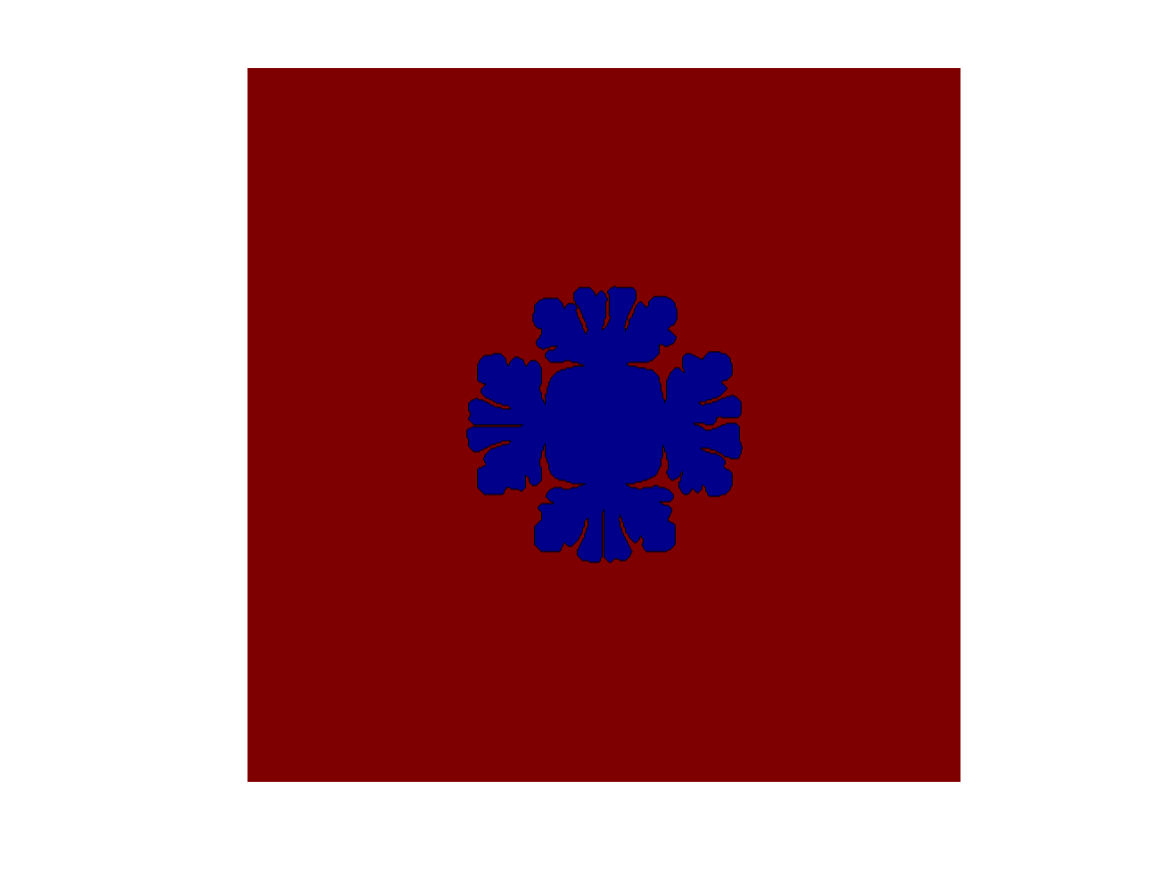}
      \end{center}
      \label{fig:s_1500}
     \end{subfigure}
     
     \bigskip
     \begin{subfigure}[ht]{0.48\textwidth}
     \caption{}
      \begin{center}
        \includegraphics[clip, trim={3cm 1cm 3cm 1.0cm},width=\textwidth]{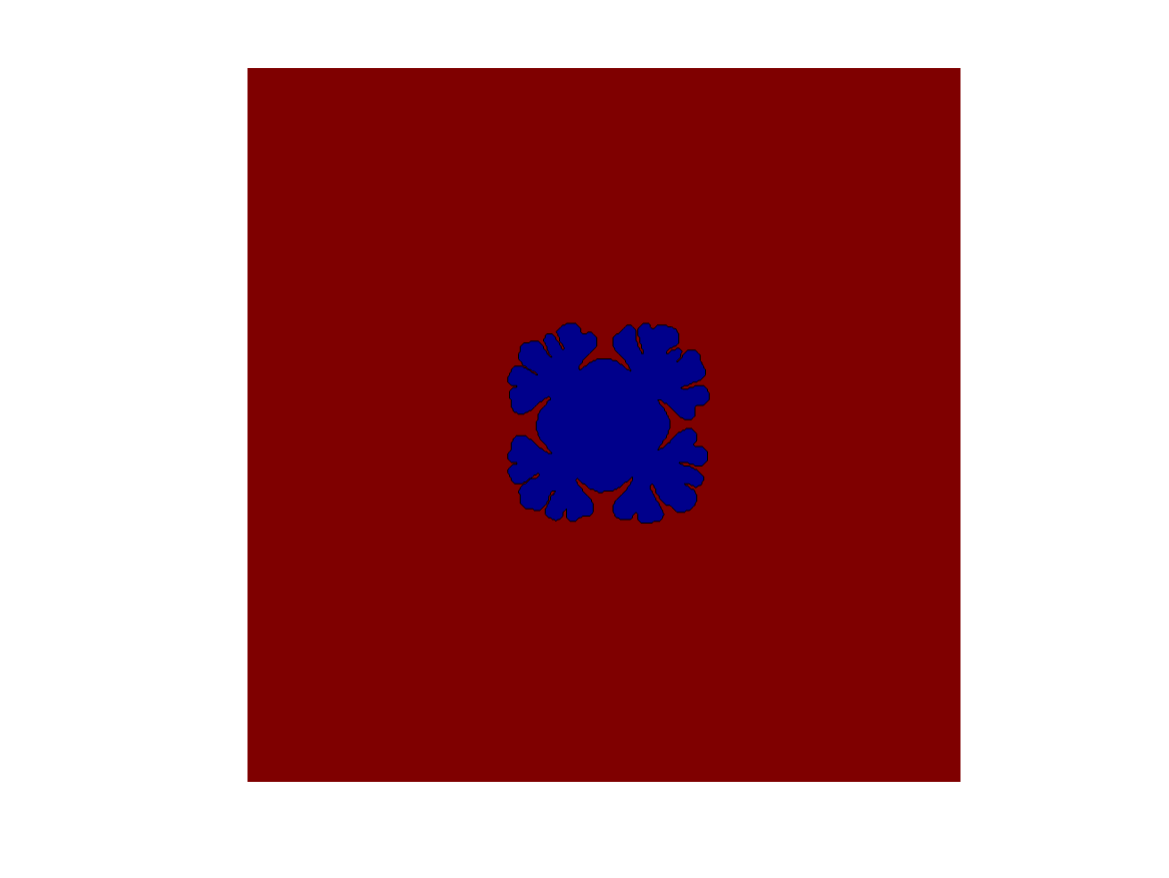}
      \end{center}
      \label{fig:rot_1000}
     \end{subfigure}
     \hfill
     \begin{subfigure}[ht]{0.48\textwidth}
      \caption{}
      \begin{center}
        \includegraphics[clip, trim={3cm 1cm 3cm 1.0cm},width=\textwidth]{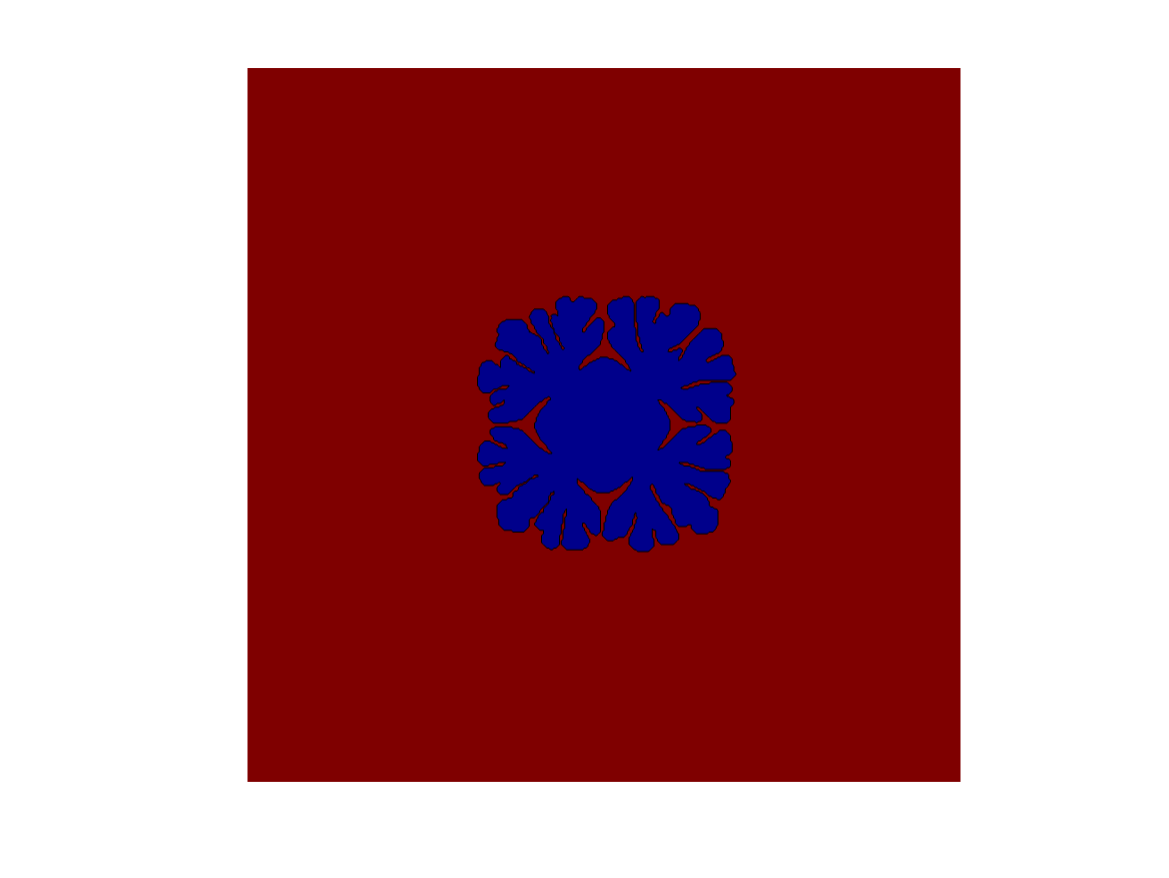}
      \end{center}
      \label{fig:rot_1500}
     \end{subfigure}
     \caption{The growth pattern of square nucleus at (I) $t = 0.01$ and (II) $t=0.015$ and rotated square nucleus at (III) $t = 0.01$ and (IV) $t=0.015$.}
     \label{fig:GOE}
\end{figure}

To evaluate  grid orientation effects (GOE), we started the simulation with a square nucleus {with} initial length $\ell_0=0.05$. The square nucleus is placed at the center of {a} unit-square domain and the initial concentration is set to $c_{\infty}=1$. The concentration at radius $r_{\infty} = 0.5$ is kept constant at $c_{\infty}$, i.e. $c(r_{\infty})=c_{\infty}$. The Damk\"{o}hler number $Da$ of the simulation is 10000. In the second simulation, the square nucleus is rotated 45$^{\circ}$ to evaluate  GOE. Figure~\ref{fig:GOE} shows snapshots of the growth pattern for the original and rotated square nucleus at $t=0.01$ and $t=0.015$, respectively. No significant GOE is observed between the two growth patterns.

\section{Conclusions}\label{sec:conclusions}
We introduced a Level Set-Immersed Boundary Method (LS-IBM) for reactive transport involving moving boundaries. The proposed method extends the GCIBM by \cite{yousefzadeh2019high} to problems with  moving boundaries. The level set method keeps track of the evolving interface, while the immersed boundary captures the reactive boundary condition. To the best of our knowledge, this is the  first method to couple level set  with the immersed boundary method for reactive transport problems involving a moving boundary.  The proposed method also ensures at least second-order accuracy in space throughout the process as well as a more accurate formulation  to propagate the interface velocity in a narrow band near the interface. Moreover,  a consistent formulation is used to propagate the velocity for both fluid and solid regions. This is different from Li \emph{et al.} \cite{li2008level} where for each subdomain a separate method is used:  the solution of a PDE for the solid subdomain and the ghost fluid method for fluid subdomain.
 The new method uses local grid points to propagate the interface velocity, and therefore is suitable for problems involving interface phenomena. Validation studies for a cylindrical reactive calcite grain and dendritic solid growth have been successfully implemented. Further improvements, such as extensions to LS methods that conserve mass~\cite{Jettestuen2021-wy, Ge2018-eq}, three dimensional geometries and inclusion of multicomponent transport, are subject of future research.
\section*{Acknowledgments}
Full support by the Department of Energy under the Early Career award DE-SC0014227 ‘Multiscale
dynamics of reactive fronts in the subsurface’ is gratefully acknowledged.

\appendix
\section{Derivation of second-order approximation of concentration at interface}
\label{sec:2ndorder_CIB}
In Section~\ref{Sec:Problem_form}, the reactive boundary condition on the solid-fluid interface is given by Equation~\eqref{bc:reactiveBC}, i.e.
\begin{align}\label{bc:reactiveBCappen}
- n_i \dfrac{\partial C}{\partial x_i} =Da C_{\Gamma}, \quad \mathbf x\in \Gamma_{\tiny{\mbox{IB}}}.
\end{align}
To obtain a second-order accurate approximation in space of Equation~\eqref{bc:reactiveBCappen}, we use a one-sided second-order finite difference scheme, i.e.
\begin{align}\label{eq:onesideFD}
    \pdv{f}{x} = \dfrac{-3f(x) + 4 f(x+\ell) - f(x+2\ell) }{2\ell},
\end{align}
where $\ell$ is the grid spacing. Applying Equation~\eqref{eq:onesideFD} and $C_{\Gamma}=C_{\tiny{\mbox{IB}}}$ to Equation~\eqref{bc:reactiveBCappen}, we obtain 
\begin{align}\label{eq:onesideApprox}
    \dfrac{-3C_{\tiny{\mbox{IB}}} + 4 C_{X'} - C_{X''}}{2\ell} =Da C_{\tiny{\mbox{IB}}}.
\end{align}
Rearranging Equation~\eqref{eq:onesideApprox}, the second-order approximation of $C_{\tiny{\mbox{IB}}}$ is obtained as
\begin{align}
    C_{\tiny{\mbox{IB}}} = \dfrac{4C_{X'}-C_{X''}}{2\ell Da  + 3},
\end{align}
as indicated in Equation~\eqref{eq:IntConcent}.

\section{Temporal discretization and pressure-velocity coupling of the Navier-Stokes equations}
\label{sec:press-vel-coupling}
\noah{One of the challenges in solving Navier-Stokes equations~\eqref{eq:NavierStokes} is related to the coupling between the pressure and velocity fields. In this study, we adopted the PISO algorithm~\cite{issa1986solution} that uses a predictor-corrector step for pressure-velocity coupling. To eliminate the time-step constraints related to convective and viscous terms in the equations, we adopted an Euler backward time integration scheme for temporal integration. In this section, we describe in detail the marching of the solution from time $t^n$ to $t^{n+1}$. 
\begin{enumerate}[(a)]
    \item \emph{Predictor Step. ($m=0$)} An intermediate velocity $u_{i}^{0}$ is implicitly calculated with the prevailing velocity $u^n$ and pressure $p^n$ at time $t^n$.
    \begin{gather}
    u_i^{0} = u_i^n + \Delta t\left( \mathcal{H}_{u}\left(u_i^{0}\right) -\frac{\partial p^n}{\partial x_i}\right), \label{eq:pred-u-star}
    \end{gather}
    where $\mathcal{H}_{u}\left(u_i^{0}\right) = \mathcal{C}_{u}\left(u_i^{0}\right) + \mathcal{D}_{u}\left(u_i^{0}\right) + \mathcal{F}_{u}\left(u_i^{0}\right)$ is the discretized representation of the convective term ($\mathcal{C}_{u}$), diffusive term ($\mathcal{D}_{u}$) and the forcing term due to the immersed boundary method ($\mathcal{F}_{u}$). 
    
    \item \emph{First Corrector Step ($m=1$).} A new velocity $u_i^{1}$ and pressure $p^{0}$ are calculated such that
    \begin{gather}
    \dfrac{ \partial u_i^{1}}{\partial x_i} = 0, \label{eq:1st-u-div} \\
    u_i^{1} = u_i^n + \Delta t\left( \mathcal{H}_{u}\left(u_i^{0}\right) -\frac{\partial p^{0}}{\partial x_i}\right). \label{eq:1st-u-star} 
    \end{gather}
    By subtracting equation~\eqref{eq:pred-u-star} from equation~\eqref{eq:1st-u-star}, we obtain
    \begin{gather}
    u_i^{1} = u_i^{0}  -\Delta{t} \frac{\partial  p_{cor}^0}{\partial x_i}, \label{eq:1st-u-update}
    \end{gather}
    where $p^{0} = p^n + p^{0}_{cor}$ and $p^{0}_{cor}$ is the correction pressure for $m=1$. By applying the divergence operator to equation~\eqref{eq:1st-u-update} and substituting equation~\eqref{eq:1st-u-div}, we obtain the Poisson equation to calculate $p^{0}_{cor}$ as
    \begin{gather}
    \pdv{}{x_i}\left(\pdv{p_{cor}^0}{x_i}\right) = \frac{1}{\Delta {t}} \pdv{u_i^{0}}{x_i}. \label{eq:1st-p-cor}
    \end{gather}
    Once $p_{cor}^0$ is computed, $u_i^1$ can be updated with equation~\eqref{eq:1st-u-update}. 
    
    \item \emph{Second Corrector Step ($m=2$)}. New velocity $u_i^{2}$ and pressure $p^{1}$ fields are computed such that 
    \begin{gather}
    \dfrac{ \partial u_i^{2}}{\partial x_i} = 0, \label{eq:2nd-u-div} \\
    u_i^{2} = u_i^n + \Delta t\left( \mathcal{H}_{u}\left(u_i^{1}\right) -\frac{\partial p^{1}}{\partial x_i}\right). \label{eq:2nd-u-star}
    \end{gather}
    By following similar procedures, we obtain the equations to update the velocity and pressure fields as
    \begin{gather}
    u_i^{2} = u_i^{1} + \Delta{t}\mathcal{H}_{u}\left(u^{\prime}_i\right) -  \Delta{t} \frac{\partial \left( p_{cor}^{1} \right)}{\partial x_i}, \\
     \pdv{\mathcal{H}_{u}\left(u^\prime_i\right)}{x_i} = \pdv{}{x_i} \left( \frac{\partial p_{cor}^{1}}{\partial x_i} \right),\\
     p^{1} = p^{0} + p_{cor}^{1},
    \end{gather}
    where $u^{\prime}_i = u_i^{1} - u_i^{0}$ is the difference between the velocity at $m=1$ and $m=0$. Following the definition in Issa \emph{et al.}~\cite{issa1986solution}, velocity $u_i^{2}$ and pressure $p^{1}$ are defined as the respective fields at time $t^{n+1}$ such that $u_i^{n+1} = u_i^{2}$ and $p^{n+1} = p^1$. In practice, more correction steps ($m > 2$) can be performed to ensure the accuracy of the solutions.
\end{enumerate}}

\section{Temporal discretization of the scalar transport and Level-Set equations}
\label{app:other-temp}
\noah{For the scalar transport equation~\eqref{eq:transport}, we adopted the same temporal discretization scheme as the Navier-Stokes equations using Backward Euler such that 
\begin{gather}
    C^{n+1} = C^n + \Delta t \mathcal{H}_{C}\left(C^{n+1}\right),
\end{gather}
where $\mathcal{H}_{C}\left(C^{n+1}\right) = \mathcal{C}_C\left(C^{n+1}\right) + \mathcal{D}_C\left(C^{n+1}\right)$ is the discretized representation of the convective and diffusive terms in the scalar transport equation.

For the level-set equation~\eqref{eq:levelset}, we adopted a TVD-RK3~\cite{shu1988efficient} for temporal discretization as 
\begin{subequations}
\begin{gather}
    \phi^1 = \phi^n + \Delta{t} \left(\mathcal{H}_{\phi} \left(\mathbf{u}^n_{\Gamma}, \phi^n \right)\right), \\
    \phi^2 = \phi^n + \frac{\Delta{t}}{4} \left(\mathcal{H}_{\phi} \left(\mathbf{u}^n_{\Gamma}, \phi^n \right) + \mathcal{H}_{\phi} \left(\mathbf{u}^n_{\Gamma}, \phi^1 \right)\right), \\
    \phi^{n+1} = \phi^n + \frac{\Delta{t}}{6} \left(\mathcal{H}_{\phi} \left(\mathbf{u}^n_{\Gamma}, \phi^n \right) + \mathcal{H}_{\phi} \left(\mathbf{u}^n_{\Gamma}, \phi^1 \right) + \mathcal{H}_{\phi} \left(\mathbf{u}^n_{\Gamma}, \phi^2 \right) \right),
\end{gather}
\end{subequations}
where $\mathcal{H}_{\phi} = -\mathbf{u}_{\Gamma}\cdot\nabla \phi$ is the discretized representation of the convective term. 
Similar temporal discretization is adopted for the reinitialization equation \eqref{eq:reinit}. The semi-discretized equations are
\begin{subequations}
    \begin{gather}
        \phi^1 = \phi^n + \Delta{t} \left(\mathcal{T}_{\phi} \left(\phi^0, \phi^n \right)\right), \\
        \phi^2 = \phi^n + \frac{\Delta{t}}{4} \left(\mathcal{T}_{\phi} \left(\phi^0, \phi^n \right) + \mathcal{T}_{\phi} \left(\phi^0, \phi^1 \right)\right), \\
        \phi^{n+1} = \phi^n + \frac{\Delta{t}}{6} \left(\mathcal{T}_{\phi} \left(\phi^0, \phi^n \right) + \mathcal{T}_{\phi} \left(\phi^0, \phi^1 \right) + \mathcal{T}_{\phi} \left(\phi^0, \phi^2 \right) \right).
    \end{gather}
\end{subequations}
where $\mathcal{T}_{\phi} = -\left(\text{sgn}(\phi^0)(1-|\nabla \phi|)\right)$ is the discretized reprenstation. }

\bibliographystyle{elsarticle-num}

\begin{thebibliography}{10}
\expandafter\ifx\csname url\endcsname\relax
  \def\url#1{\texttt{#1}}\fi
\expandafter\ifx\csname urlprefix\endcsname\relax\def\urlprefix{URL }\fi
\expandafter\ifx\csname href\endcsname\relax
  \def\href#1#2{#2} \def\path#1{#1}\fi

\bibitem{morse2007calcium}
J.~W. Morse, R.~S. Arvidson, A.~L{\"u}ttge, Calcium carbonate formation and
  dissolution, Chemical reviews 107~(2) (2007) 342--381.

\bibitem{lichtner2018reactive}
P.~C. Lichtner, C.~I. Steefel, E.~H. Oelkers, Reactive transport in porous
  media, Vol.~34, Walter de Gruyter GmbH \& Co KG, 2018.

\bibitem{yousefzadeh2017physics}
M.~Yousefzadeh, I.~Battiato, Physics-based hybrid method for multiscale
  transport in porous media, Journal of Computational Physics 344 (2017)
  320--338.

\bibitem{ryan2013computational}
E.~M. Ryan, K.~Ferris, A.~Tartakovsky, M.~Khaleel, Computational modeling of
  transport limitations in li-air batteries, ECS Transactions 45~(29) (2013)
  123--136.

\bibitem{tan2016computational}
J.~Tan, E.~M. Ryan, Computational study of electro-convection effects on
  dendrite growth in batteries, Journal of Power Sources 323 (2016) 67--77.

\bibitem{pournik2014effect}
M.~Pournik*, D.~Tripathi, Effect of acid on productivity of fractured shale
  reservoirs, in: Unconventional Resources Technology Conference, Denver,
  Colorado, 25-27 August 2014, Society of Exploration Geophysicists, American
  Association of Petroleum~…, 2014, pp. 1811--1823.

\bibitem{teklu2017experimental}
T.~W. Teklu, H.~H. Abass, R.~Hanashmooni, J.~C. Carratu, M.~Ermila,
  Experimental investigation of acid imbibition on matrix and fractured
  carbonate rich shales, Journal of Natural Gas Science and Engineering 45
  (2017) 706--725.

\bibitem{pournik2019productivity}
M.~Pournik, D.~Tripathi, I.~El-Monier, Productivity of hydraulically fractured
  calcite rich shale reservoirs enhanced by acid, Journal of Petroleum
  Engineering \& Technology 6~(1) (2019) 54--68.

\bibitem{deng2017alteration}
H.~Deng, M.~Voltolini, S.~Molins, C.~Steefel, D.~DePaolo, J.~Ajo-Franklin,
  L.~Yang, Alteration and erosion of rock matrix bordering a carbonate-rich
  shale fracture, Environmental science \& technology 51~(15) (2017)
  8861--8868.

\bibitem{rathnaweera2016experimental}
T.~Rathnaweera, P.~Ranjith, M.~Perera, Experimental investigation of
  geochemical and mineralogical effects of co 2 sequestration on flow
  characteristics of reservoir rock in deep saline aquifers, Scientific reports
  6 (2016) 19362.

\bibitem{xiao2009effects}
Y.~Xiao, T.~Xu, K.~Pruess, The effects of gas-fluid-rock interactions on co2
  injection and storage: insights from reactive transport modeling, Energy
  Procedia 1~(1) (2009) 1783--1790.

\bibitem{xu2003reactive}
T.~Xu, J.~A. Apps, K.~Pruess, Reactive geochemical transport simulation to
  study mineral trapping for co2 disposal in deep arenaceous formations,
  Journal of Geophysical Research: Solid Earth 108~(B2).

\bibitem{huerta2016reactive}
N.~J. Huerta, M.~A. Hesse, S.~L. Bryant, B.~R. Strazisar, C.~Lopano, Reactive
  transport of co2-saturated water in a cement fracture: Application to
  wellbore leakage during geologic co2 storage, International Journal of
  Greenhouse Gas Control 44 (2016) 276--289.

\bibitem{wuana2011heavy}
R.~A. Wuana, F.~E. Okieimen, Heavy metals in contaminated soils: a review of
  sources, chemistry, risks and best available strategies for remediation, Isrn
  Ecology 2011.

\bibitem{guo2018stability}
F.~Guo, C.~Ding, Z.~Zhou, G.~Huang, X.~Wang, Stability of immobilization
  remediation of several amendments on cadmium contaminated soils as affected
  by simulated soil acidification, Ecotoxicology and environmental safety 161
  (2018) 164--172.

\bibitem{soulaine2017mineral}
C.~Soulaine, S.~Roman, A.~Kovscek, H.~A. Tchelepi, Mineral dissolution and
  wormholing from a pore-scale perspective, Journal of Fluid Mechanics 827
  (2017) 457--483.

\bibitem{tartakovsky2007smoothed}
A.~M. Tartakovsky, P.~Meakin, T.~D. Scheibe, B.~D. Wood, A smoothed particle
  hydrodynamics model for reactive transport and mineral precipitation in
  porous and fractured porous media, Water resources research 43~(5).

\bibitem{huber2014new}
C.~Huber, B.~Shafei, A.~Parmigiani, A new pore-scale model for linear and
  non-linear heterogeneous dissolution and precipitation, Geochimica et
  Cosmochimica Acta 124 (2014) 109--130.

\bibitem{parmigiani2011pore}
A.~Parmigiani, C.~Huber, O.~Bachmann, B.~Chopard, Pore-scale mass and reactant
  transport in multiphase porous media flows, Journal of Fluid Mechanics 686
  (2011) 40--76.

\bibitem{prasianakis2017deciphering}
N.~Prasianakis, E.~Curti, G.~Kosakowski, J.~Poonoosamy, S.~Churakov,
  Deciphering pore-level precipitation mechanisms, Scientific reports 7~(1)
  (2017) 13765.

\bibitem{yousefzadeh2019high}
M.~Yousefzadeh, I.~Battiato, High order ghost-cell immersed boundary method for
  generalized boundary conditions, International Journal of Heat and Mass
  Transfer 137 (2019) 585--598.

\bibitem{Yao2022-aq}
Y.~Yao, E.~Biegert, B.~Vowinckel, T.~K{\"o}llner, E.~Meiburg, S.~Balachandar,
  C.~S. Criddle, O.~B. Fringer, {Particle‐resolved simulations of four‐way
  coupled, polydispersed, particle‐laden flows}, Int. J. Numer. Methods
  Fluids\href {http://dx.doi.org/10.1002/fld.5128}
  {\path{doi:10.1002/fld.5128}}.

\bibitem{molins2015reactive}
S.~Molins, Reactive interfaces in direct numerical simulation of pore-scale
  processes, Reviews in Mineralogy and Geochemistry 80~(1) (2015) 461--481.

\bibitem{unverdi1992front}
S.~O. Unverdi, G.~Tryggvason, A front-tracking method for viscous,
  incompressible, multi-fluid flows, Journal of computational physics 100~(1)
  (1992) 25--37.

\bibitem{juric1996front}
D.~Juric, G.~Tryggvason, A front-tracking method for dendritic solidification,
  Journal of computational physics 123~(1) (1996) 127--148.

\bibitem{osher1988fronts}
S.~Osher, J.~A. Sethian, Fronts propagating with curvature-dependent speed:
  algorithms based on hamilton-jacobi formulations, Journal of computational
  physics 79~(1) (1988) 12--49.

\bibitem{osher2001level}
S.~Osher, R.~P. Fedkiw, Level set methods: an overview and some recent results,
  Journal of Computational physics 169~(2) (2001) 463--502.

\bibitem{hirt1981volume}
C.~W. Hirt, B.~D. Nichols, Volume of fluid (vof) method for the dynamics of
  free boundaries, Journal of computational physics 39~(1) (1981) 201--225.

\bibitem{xu2011phase}
Z.~Xu, P.~Meakin, Phase-field modeling of two-dimensional solute
  precipitation/dissolution: Solid fingers and diffusion-limited precipitation,
  The Journal of chemical physics 134~(4) (2011) 044137.

\bibitem{xu2012phase}
Z.~Xu, H.~Huang, X.~Li, P.~Meakin, Phase field and level set methods for
  modeling solute precipitation and/or dissolution, Computer Physics
  Communications 183~(1) (2012) 15--19.

\bibitem{osher2004level}
S.~Osher, R.~Fedkiw, K.~Piechor, Level set methods and dynamic implicit
  surfaces, Appl. Mech. Rev. 57~(3) (2004) B15--B15.

\bibitem{fedkiw1999non}
R.~P. Fedkiw, T.~Aslam, B.~Merriman, S.~Osher, A non-oscillatory eulerian
  approach to interfaces in multimaterial flows (the ghost fluid method),
  Journal of computational physics 152~(2) (1999) 457--492.

\bibitem{li2008level}
X.~Li, H.~Huang, P.~Meakin, Level set simulation of coupled advection-diffusion
  and pore structure evolution due to mineral precipitation in porous media,
  Water Resources Research 44~(12) (2008) 1--17.

\bibitem{li2010three}
X.~Li, H.~Huang, P.~Meakin, A three-dimensional level set simulation of coupled
  reactive transport and precipitation/dissolution, International Journal of
  Heat and Mass Transfer 53~(13-14) (2010) 2908--2923.

\bibitem{molins2017mineralogical}
S.~Molins, D.~Trebotich, G.~H. Miller, C.~I. Steefel, Mineralogical and
  transport controls on the evolution of porous media texture using direct
  numerical simulation, Water Resources Research 53~(5) (2017) 3645--3661.

\bibitem{molins2020simulation}
S.~Molins, C.~Soulaine, N.~I. Prasianakis, A.~Abbasi, P.~Poncet, A.~J. Ladd,
  V.~Starchenko, S.~Roman, D.~Trebotich, H.~A. Tchelepi, et~al., Simulation of
  mineral dissolution at the pore scale with evolving fluid-solid interfaces:
  Review of approaches and benchmark problem set, Computational Geosciences
  (2020) 1--34.

\bibitem{chai2019finite}
M.~Chai, K.~Luo, C.~Shao, H.~Wang, J.~Fan, A finite difference discretization
  method for heat and mass transfer with robin boundary conditions on irregular
  domains, Journal of Computational Physics (2019) 108890.

\bibitem{trebotich2015adaptive}
D.~Trebotich, D.~Graves, An adaptive finite volume method for the
  incompressible navier--stokes equations in complex geometries, Communications
  in Applied Mathematics and Computational Science 10~(1) (2015) 43--82.

\bibitem{trebotich2014high}
D.~Trebotich, M.~F. Adams, S.~Molins, C.~I. Steefel, C.~Shen, High-resolution
  simulation of pore-scale reactive transport processes associated with carbon
  sequestration, Computing in Science \& Engineering 16~(6) (2014) 22--31.

\bibitem{luo2016ghost}
K.~Luo, Z.~Zhuang, J.~Fan, N.~E.~L. Haugen, A ghost-cell immersed boundary
  method for simulations of heat transfer in compressible flows under different
  boundary conditions, International Journal of Heat and Mass Transfer 92
  (2016) 708--717.

\bibitem{yousefzadeh2020numerical}
M.~Yousefzadeh, Numerical simulation of fluid - mineral interaction and
  reactive transport in porous and fractured media, Stanford University, 2020.

\bibitem{Jettestuen2021-wy}
E.~Jettestuen, H.~A. Friis, J.~O. Helland, {A locally conservative multiphase
  level set method for capillary-controlled displacements in porous media}, J.
  Comput. Phys. 428 (2021) 109965.
\newblock \href {http://dx.doi.org/10.1016/j.jcp.2020.109965}
  {\path{doi:10.1016/j.jcp.2020.109965}}.

\bibitem{Ge2018-eq}
Z.~Ge, J.-C. Loiseau, O.~Tammisola, L.~Brandt, {An efficient mass-preserving
  interface-correction level set/ghost fluid method for droplet suspensions
  under depletion forces}, J. Comput. Phys. 353 (2018) 435--459.
\newblock \href {http://dx.doi.org/10.1016/j.jcp.2017.10.046}
  {\path{doi:10.1016/j.jcp.2017.10.046}}.

\bibitem{sussman1994level}
M.~Sussman, P.~Smereka, S.~Osher, A level set approach for computing solutions
  to incompressible two-phase flow, Journal of Computational physics 114~(1)
  (1994) 146--159.

\bibitem{ferziger2012computational}
J.~H. Ferziger, M.~Peric, Computational methods for fluid dynamics, Springer
  Science \& Business Media, 2012.

\bibitem{versteeg2007introduction}
H.~K. Versteeg, W.~Malalasekera, An introduction to computational fluid
  dynamics: the finite volume method, Pearson Education, 2007.

\bibitem{harlow1965numerical}
F.~H. Harlow, J.~E. Welch, Numerical calculation of time-dependent viscous
  incompressible flow of fluid with free surface, The physics of fluids 8~(12)
  (1965) 2182--2189.

\bibitem{issa1986solution}
R.~I. Issa, Solution of the implicitly discretised fluid flow equations by
  operator-splitting, Journal of computational physics 62~(1) (1986) 40--65.

\bibitem{leonard1979stable}
B.~P. Leonard, A stable and accurate convective modelling procedure based on
  quadratic upstream interpolation, Computer methods in applied mechanics and
  engineering 19~(1) (1979) 59--98.

\bibitem{hayase1992consistently}
T.~Hayase, J.~Humphrey, R.~Greif, A consistently formulated quick scheme for
  fast and stable convergence using finite-volume iterative calculation
  procedures, J. Comput. Phys. 98~(1) (1992) 108--118.

\bibitem{liu1994weighted}
X.-D. Liu, S.~Osher, T.~Chan, Weighted essentially non-oscillatory schemes,
  Journal of computational physics 115~(1) (1994) 200--212.

\bibitem{shu1988efficient}
C.-W. Shu, S.~Osher, Efficient implementation of essentially non-oscillatory
  shock-capturing schemes, Journal of computational physics 77~(2) (1988)
  439--471.

\bibitem{gibou2018review}
F.~Gibou, R.~Fedkiw, S.~Osher, A review of level-set methods and some recent
  applications, Journal of Computational Physics 353 (2018) 82--109.

\bibitem{Seo2011-vq}
J.~H. Seo, R.~Mittal, {A Sharp-Interface Immersed Boundary Method with Improved
  Mass Conservation and Reduced Spurious Pressure Oscillations}, J. Comput.
  Phys. 230~(19) (2011) 7347--7363.
\newblock \href {http://dx.doi.org/10.1016/j.jcp.2011.06.003}
  {\path{doi:10.1016/j.jcp.2011.06.003}}.

\bibitem{aslam2004partial}
T.~D. Aslam, A partial differential equation approach to multidimensional
  extrapolation, Journal of Computational Physics 193~(1) (2004) 349--355.

\bibitem{Ahn2007-ig}
H.~T. Ahn, M.~Shashkov, {Multi-material interface reconstruction on generalized
  polyhedral meshes}, J. Comput. Phys. 226~(2) (2007) 2096--2132.
\newblock \href {http://dx.doi.org/10.1016/j.jcp.2007.06.033}
  {\path{doi:10.1016/j.jcp.2007.06.033}}.

\bibitem{Molins2017-dx}
S.~Molins, D.~Trebotich, G.~H. Miller, C.~I. Steefel, {Mineralogical and
  transport controls on the evolution of porous media texture using direct
  numerical simulation}, Water Resour. Res. 53~(5) (2017) 3645--3661.
\newblock \href {http://dx.doi.org/10.1002/2016WR020323}
  {\path{doi:10.1002/2016WR020323}}.

\bibitem{Molins2014-pv}
S.~Molins, D.~Trebotich, L.~Yang, J.~B. Ajo-Franklin, T.~J. Ligocki, C.~Shen,
  C.~I. Steefel, {Pore-scale controls on calcite dissolution rates from
  flow-through laboratory and numerical experiments}, Environ. Sci. Technol.
  48~(13) (2014) 7453--7460.
\newblock \href {http://dx.doi.org/10.1021/es5013438}
  {\path{doi:10.1021/es5013438}}.

\bibitem{Molins2012-rk}
S.~Molins, D.~Trebotich, C.~I. Steefel, C.~Shen, {An investigation of the
  effect of pore scale flow on average geochemical reaction rates using direct
  numerical simulation}, Water Resour. Res. 48~(3).
\newblock \href {http://dx.doi.org/10.1029/2011wr011404}
  {\path{doi:10.1029/2011wr011404}}.

\bibitem{Soulaine2016-rd}
C.~Soulaine, H.~A. Tchelepi, {Micro-continuum Approach for Pore-Scale
  Simulation of Subsurface Processes}, Transp. Porous Media 113~(3) (2016)
  431--456.
\newblock \href {http://dx.doi.org/10.1007/s11242-016-0701-3}
  {\path{doi:10.1007/s11242-016-0701-3}}.

\bibitem{Yoon2012-oa}
H.~Yoon, A.~J. Valocchi, C.~J. Werth, T.~Dewers, {Pore-scale simulation of
  mixing-induced calcium carbonate precipitation and dissolution in a
  microfluidic pore network}, Water Resour. Res. 48~(2).
\newblock \href {http://dx.doi.org/10.1029/2011wr011192}
  {\path{doi:10.1029/2011wr011192}}.

\bibitem{Gazzola2011-rx}
M.~Gazzola, P.~Chatelain, W.~M. van Rees, P.~Koumoutsakos, {Simulations of
  single and multiple swimmers with non-divergence free deforming geometries},
  J. Comput. Phys. 230~(19) (2011) 7093--7114.
\newblock \href {http://dx.doi.org/10.1016/j.jcp.2011.04.025}
  {\path{doi:10.1016/j.jcp.2011.04.025}}.

\bibitem{Chatelin2016-pl}
R.~Chatelin, D.~Sanchez, P.~Poncet, {Analysis of the penalized 3D variable
  viscosity stokes equations coupled to diffusion and transport}, Esaim Math.
  Model. Numer. Anal. 50~(2) (2016) 565--591.
\newblock \href {http://dx.doi.org/10.1051/m2an/2015056}
  {\path{doi:10.1051/m2an/2015056}}.

\bibitem{Chatelin2013-aj}
R.~Chatelin, P.~Poncet, {A Hybrid Grid-Particle Method for Moving Bodies in 3D
  Stokes Flow with Variable Viscosity}, SIAM J. Sci. Comput. 35~(4) (2013)
  B925--B949.
\newblock \href {http://dx.doi.org/10.1137/120892921}
  {\path{doi:10.1137/120892921}}.

\bibitem{Starchenko2016-eb}
V.~Starchenko, C.~J. Marra, A.~J.~C. Ladd, {Three-dimensional simulations of
  fracture dissolution}, J. Geophys. Res. [Solid Earth] 121~(9) (2016)
  6421--6444.
\newblock \href {http://dx.doi.org/10.1002/2016jb013321}
  {\path{doi:10.1002/2016jb013321}}.

\bibitem{meakin1998fractals}
P.~Meakin, Fractals, scaling and growth far from equilibrium, Vol.~5, Cambridge
  university press, 1998.

\bibitem{meakin1984monte}
P.~Meakin, J.~Deutch, Monte carlo simulation of diffusion controlled colloid
  growth rates in two and three dimensions, The Journal of chemical physics
  80~(5) (1984) 2115--2122.

\bibitem{kang2004lattice}
Q.~Kang, D.~Zhang, P.~C. Lichtner, I.~N. Tsimpanogiannis, Lattice boltzmann
  model for crystal growth from supersaturated solution, Geophysical Research
  Letters 31~(21).

\bibitem{tartakovsky2007simulations}
A.~M. Tartakovsky, P.~Meakin, T.~D. Scheibe, R.~M.~E. West, Simulations of
  reactive transport and precipitation with smoothed particle hydrodynamics,
  Journal of Computational Physics 222~(2) (2007) 654--672.

\bibitem{mullins1964stability}
W.~W. Mullins, R.~Sekerka, Stability of a planar interface during
  solidification of a dilute binary alloy, Journal of applied physics 35~(2)
  (1964) 444--451.

\end{thebibliography}

\end{document}